\documentclass{intech}
\usepackage{latexsym}
\usepackage{graphicx}
\usepackage{graphics}
\usepackage{amsmath}
\usepackage{amsfonts}

\booktitle{Will-be-set-by-IN-TECH}%

\chaptertitle{A Statistical Derivation of Non-Relativistic Quantum Theory}

\authors{Ulf Klein}
\affiliation{Universitity of Linz, Institute
for Theoretical Physics}
\country{Austria}

\begin{document}

\maketitle

\section{Introduction }
\label{sec:Introduction}
Quantum theory (QT) may either be defined by a set of axioms or otherwise be 'derived' from
classical physics by using certain assumptions. Today, QT is frequently identified with a set
of axioms defining a Hilbert space structure. This mathematical structure has been created
(by von Neumann) by abstraction from the linear solution space of the central equation of
QT, the Schr{\"o}dinger equation.  Thus, deriving Schr{\"o}dinger's equation is basically
the same as deriving QT. To derive the most general version of the time-dependent
Schr{\"o}dinger equation, describing $N$ particles with spin in an external gauge field,
means to derive essentially the whole of non-relativistic QT.

The second way of proceeding is sometimes called 'quantization'. In the standard
(canonical) quantization method one starts from a classical Hamiltonian whose basic
variables are then 'transformed', by means of well-known correspondence rules,
\begin{equation}
  \label{eq:DTR43QUPR}
p\to\frac{\hbar}{\imath}\frac{\mathrm{d}}{\mathrm{d}x},\;\;\;
E\to-\frac{\hbar}{\imath}\frac{\mathrm{d}}{\mathrm{d}t}
\mbox{,}
\end{equation}
into operators. Then, all relevant classical observables may be rewritten as operators
acting on states of a Hilbert space etc; the details of the 'derivation' of Schr{\"o}dinger's
equation along this lines may be found in many textbooks. There are formal problems
with this approach  which have been identified many years ago, and can be expressed e.g. in
terms of Groenewold's theorem, see~\cite{groenewold:principles},~\cite{gotay:groenewold}. Even
more seriously, there is no satisfactory \emph{explanation} for this 'metamorphosis' of
observables into operators. This quantization method (as well as several other mathematically
more sophisticated versions of it) is just a \emph{recipe} or, depending on one's taste,
"black magic", \cite{hall:exact_uncertainty}. Note that the enormous success of
this recipe in various contexts - including field quantization - is no substitute for an explanation.

The choice of  a particular quantization procedure will be strongly influenced by the preferred
interpretation of the quantum theoretical formalism. If QT is interpreted as a theory describing
individual events, then the Hamiltonian of classical mechanics becomes a natural starting point.
This 'individuality assumption' is an essential part of the dominating 'conventional', or
'Copenhagen', interpretation (CI) of QT. It is well-known, that QT becomes a source of mysteries
and paradoxes\footnote{I cannot report here a list, all the less a description, of all the quantum
mechanical paradoxes found in the last eighty years.} whenever it is interpreted in the sense of CI,
as a (complete) theory for individual events. Thus, the canonical quantization method and the CI
are similar in two respects: both rely heavily on the concept of individual particles and both are
rather mysterious.

This situation confronts us with a fundamental alternative. Should we accept the mysteries
and paradoxes as inherent attributes of reality or should we not, instead, critically reconsider
our assumptions, in particular the 'individuality assumption'. As a matter of fact, the dynamical
numerical output of quantum mechanics consists of \emph{probabilities}. A probability is a
"deterministic" prediction which can be verified in a statistical sense only, i.e. by
performing experiments on a large number of identically prepared individual
systems, see~\cite{belinfante:individual},~\cite{margenau:measurements}.
Therefore, the very structure of QT tells us that it is a theory about statistical
ensembles only, see~\cite{ballentine:statistical}. If dogmatic or philosophical reasons 'force' us
to interpret QT as a theory about individual events, we have to create complicated
intellectual constructs, which are not part of the physical formalism, and lead to unsolved
problems and contradictions.

The present author believes, like
several other physicists [see e.g.~\cite{kemble:principles1,einstein:physics_reality,
margenau:quantum-mechanical,blokhintsev:quantum,ballentine:statistical,belinfante:measurements,
ross-bonney:dice,young:quantum,newton:probability,pippard:interpretation,tschudi:statistical,
toyozawa:measurement,krueger:epr_debate,ali:ensemble}] that QT is a purely statistical theory
whose predictions can only be used to describe the behavior of statistical ensembles and not of
individual particles. This statistical interpretation (SI) of QT eliminates all mysteries and
paradoxes - and this shows that the mysteries and paradoxes are not part of QT itself but rather
the result of a particular (mis)interpretation of QT. In view of the similarity discussed above,  we
adopt the statistical point of view, not only for the interpretation of QT itself, but also in our search
for \emph{quantization conditions}. The general strategy is to find a new set of (as) simple (as possible)
statistical assumptions which can be understood in physical terms and imply QT. Such an approach
would also provide an explanation for the correspondence rules~\eqref{eq:DTR43QUPR}.

The present paper belongs to a series of works aimed at such an explanation. Quite generally,
the present work continues a long tradition of attempts, see~\cite{schrodinger:quantisierung_I,
motz:quantization,schiller:quasiclassical,rosen:classical_quantum,frieden:fisher_basis,
lee.zhu:principle,hall.reginatto:quantum_heisenberg,frieden:sciencefisher}, to characterize QT by mathematical relations which can be understood
in physical terms\footnote{The listing given here is far from complete.} (in contrast to
the axiomatic approach). More specifically, it continues previous attempts to derive
Schr{\"o}dinger's equation with the help of statistical
concepts, see~\cite{hall.reginatto:schroedinger},~\cite{reginatto:derivation,
syska:fisher},~\cite{klein:schroedingers}. These works, being quite different in detail, share
the common feature that a statistical ensemble and not a particle Hamiltonian is used
as a starting point for quantization.  Finally, in a previous work,~\cite{klein:statistical}, of the
present author an attempt has been undertaken to construct a complete statistical approach to
QT with the help of a small number of very simple (statistical) assumptions. This work will be
referred to as I. The present paper is a continuation and extension of I.

The quantization method reported in I is based on the following general ideas:
\begin{itemize}
\item QT should be a probabilistic theory in configuration space (not in phase space).
\item QT should fullfil abstract versions of (i) a conservation law for probability (continuity
equation) , and (ii) Ehrenfest's theorem. Such relations hold in all statistical theories no
matter whether quantum or classical.
\item There are no laws for particle trajectories in QT anymore. This arbitrariness, which
represents a crucial difference between QT and classical statistics, should be handled by a
statistical principle analogous to the principle of maximal entropy in classical statistics.
\end{itemize}

These general ideas lead to the mathematical assumptions which
represent the basis for the treatment reported in I. This work was restricted to a
one-dimensional configuration space (a single particle ensemble with a single
spatial degree of freedom). The present work generalizes the treatment of I
to a $3N-$dimensional configuration space ( ensembles representing an arbitrary
number $N$ of particles allowed to move in three-dimensional space), gauge-coupling,
and spin. In a first step the generalization to three spatial dimensions is performed; the properly
generalized basic relations are reported in section~\ref{sec:calcwpt}. This section
contains also a review of the fundamental ideas.

In section~\ref{sec:gaugecoupling} we make use of a mathematical freedom, which is
already contained in our basic assumptions, namely the multi-valuedness of the
variable $S$. This leads to the appearance of potentials in statistical relations replacing
the local forces of single-event (mechanical) theories. The mysterious non-local action
of the vector potential (in effects of the Aharonov-Bohm type) is explained
as a consequence of the statistical nature of QT.  In section~\ref{sec:stat-constr-macr} we
discuss a related question: Which constraints on admissible forces exist for the present
class of statistical theories ? The answer is that only macroscopic (elementary) forces
of the form of the Lorentz force can occur in nature, because only these survive the
transition to QT . These forces are statistically represented by potentials, i.e. by the familiar
gauge coupling terms in matter field equations. The present statistical approach
provides a natural explanation for the long-standing question why potentials
play an indispensable role in the field equations of physics.

In section~\ref{sec:fisher-information} it is shown that among all
statistical theories only the time-dependent Schr\"odinger equation
follows the logical requirement of maximal disorder or minimal Fisher
information. Spin one-half is introduced, in section~\ref{sec:spin},
as the property of a statistical ensemble to respond to an external gauge
field in two different ways. A generalized calculation, reported
in sections~\ref{sec:spin} and~\ref{sec:spin-fish-inform}, leads to
Pauli's (single-particle) equation.  In section~\ref{sec:spin-as-gauge} an
alternative derivation, following \cite{arunsalam:hamiltonians},
and \cite{gould:intrinsic} is reported, which is particularly convenient
for the generalization to arbitrary $N$. The latter is performed in
section~\ref{sec:final-step-to}, which completes our statistical derivation
of non-relativistic QT.

In section~\ref{sec:classical-limit} the classical limit of QT is  studied and
it is stressed that the classical limit of QT is not classical mechanics but a
classical statistical theory. In section~\ref{sec:discussion}
various questions related to the present approach, including the
role of potentials and the interpretation of QT, are discussed.
The final section~\ref{sec:concludingremarks} contains a short
summary and mentions a possible direction for future research.

\section{Basic equations for a class of statistical theories}
\label{sec:calcwpt}
In I three different types of theories have been defined which differ from each
other with regard to the role of probability. We give a short review of the defining
properties and supply some additional comments characterizing these theories.

The dogma underlying \emph{theories of type 1} is determinism with regard to
single events; probability does not play any role. If nature behaves according
to this dogma, then measurements on identically prepared individual systems
yield identical results. Classical mechanics is obviously such a deterministic
type 1 theory. We shall use below (as a 'template' for the dynamics of our statistical
theories) the following version of Newton's law, where the particle momentum
$p_k(t)$ plays the role of a second dynamic variable besides the spatial coordinate $x_k(t)$:
\begin{equation}
\label{eq:TVONLIUSCHL}
\frac{\mathrm{d}}{\mathrm{d}t} x_k(t)  =  \frac{p_k(t)}{m},
\hspace{0.5cm}
\frac{\mathrm{d}}{\mathrm{d}t} p_k(t)  =   F_k(x,\,p,\,t)
\mbox{.}
\end{equation}
In classical mechanics there is no restriction as regards the admissible
forces. Thus,  $F_k$ is an arbitrary function of $x_k,\,p_k,\,t$; it is,
in particular, not required that it be derivable from a potential. Note
that Eqs.~\eqref{eq:TVONLIUSCHL} do \emph{not} hold in the present
theory; these relations are just used to establish a correspondence
between classical mechanics and associated statistical theories.

Experimental data from atomic systems, recorded since the beginning of
the last century, indicate that nature does not behave according to this
single-event deterministic dogma. A simple but somewhat unfamiliar idea
is, to construct a theory which is deterministic only in a statistical sense.
This means that measurements on identically prepared individual systems
do not yield identical results (no determinism with regard to single events) but
repeated measurements on ensembles [consisting each time of a large
(infinite) number of measurements on individual systems] yield identical
results. In this case we have 'determinism' with regard to ensembles
(expectation values, or probabilities).

Note that such a theory is far from chaotic even if our
macroscopic anticipation of (single-event) determinism is not satisfied.
Note also that there is no reason to assume that such a statistical
theory for microscopic events is incompatible with macroscopic
determinism. It is a frequently  observed (but not always completely
understood) phenomenon in nature that systems with many (microscopic)
degrees of freedom can be described by a much smaller number of variables.
During this process of elimination of variables the details of the
corresponding microscopic theory for the individual constituents are
generally lost. In other words, there is no reason to assume that
a fundamental statistical law for individual atoms and a
deterministic law for a piece of matter consisting of, say,
$10^{23}$ atoms should not be compatible with each other. This way
of characterizing the relation between two physical theories
is completely different from the common reductionistic point of view.
Convincing arguments in favor of the former may, however, be
found in~\cite{anderson:more},~\cite{laughlin:different}.

As discussed in I two types (referred to as type 2 and type 3) of
indeterministic theories may be identified. In \emph{type 2 theories}
laws for individual particles exist (roughly speaking the individuality
of particles remains intact) but the initial values are unknown and
are described by probabilities only. An example for such a
(classical-statistical) type 2 theory is statistical thermodynamics.
On the other hand, in \emph{type 3 theories} the amount of uncertainty is
still greater, insofar as no dynamic laws for individual particles
exist any more. A possible candidate for this 'extreme' type
of indeterministic theory is quantum mechanics.

The method used in I to construct statistical theories was based on the
following three assumptions,
\begin{itemize}
\item A local conservation law of probability with a particular
form of the probability current.
\item Two differential equations which are similar in structure to the canonical
equations~\eqref{eq:TVONLIUSCHL} but with observables replaced by expectation
values.
\item A differential version (minimal Fisher information) of the statistical principle
of maximal disorder.
\end{itemize}
These (properly generalized) assumptions represent also the formal basis of
the present work. The first and second of these cover type 2 as well as type 3 theories,
while it will be shown that the third - the requirement of maximal disorder - does only
hold for a single type 3 theory, namely quantum mechanics. In this sense quantum
mechanics may be considered as the \emph{most reasonable} theory among all
statistical theories defined by the first two assumptions. There is obviously an analogy
between quantum mechanics and the principle of minimal Fisher information
on the one hand and classical statistical mechanics and the principle of maximal
entropy on the other hand; both theories are realizations of the principle of
maximal disorder.

Let us now generalize the basic equations of I (see section 3 of I) with respect to the number
of spatial dimensions and with respect to gauge freedom. The continuity equation takes the form
\begin{equation}
  \label{eq:CONT3DSCHL}
\frac{\partial \rho(x,t)}{\partial t}+
\frac{\partial}{\partial x_k} \frac{\rho(x,t)}{m}\frac{\partial
\tilde{S}(x,t)}{\partial x_k}=0
\mbox{.}
\end{equation}
We use the summation convention, indices $i,\,k,...$ run from $1$ to $3$  and are omitted if
the corresponding variable occurs in the argument of a function. The existence of a local
conservation law for the probability density  $\rho(x,t)$ is a necessity for a probabilistic theory.
The same is true for the fact that the probability current takes the
form $j_k(x,t) = \rho(x,t) \tilde{p}_k(x,t)/m$,
where $\tilde{p}_k(x,t)$ is the $k-$th component of the momentum probability density.
The only non-trivial assumption contained in~\eqref{eq:CONT3DSCHL}, is the fact
that $\tilde{p}_k(x,t)$ takes the form of the gradient,
\begin{equation}
  \label{eq:DEFMOMSCHL}
\tilde{p}_k(x,\,t)=\frac{\partial \tilde{S}(x,\,t)}{\partial x_k}
\mbox{,}
\end{equation}
of a function $\tilde{S}(x,t)$. In order to gain a feeling for the physical meaning
of~\eqref{eq:DEFMOMSCHL} we could refer to the fact that a similar relation
may be found in the Hamilton-Jacobi formulation of classical mechanics~\cite{synge:classical};
alternatively we could also refer to the fact that this condition characterizes
'irrotational flow' in fluid mechanics. Relation~\eqref{eq:DEFMOMSCHL} could also
be justified by means of the principle of simplicity; a gradient is the simplest way to
represent a vector field, because it can be derived from a single scalar function.

In contrast to I we allow now for \emph{multi-valued} functions $\tilde{S}(x,t)$.
At first sight this seems strange since a multi-valued quantity cannot be an observable
and should, consequently, not appear in equations bearing a physical meaning. However, only
derivatives of $\tilde{S}(x,t)$ occur in our basic equations. Thus, this freedom is possible without
any additional postulate; we just have to require that
\begin{equation}
  \label{eq:TGT7RMT}
\tilde{S}(x,t)\;\;\;\mbox{multi-valued},
\;\;\;\;\;
\frac{\partial \tilde{S}}{\partial t},
\frac{\partial \tilde{S}}{\partial x_{k}}
\;\;\;\mbox{single-valued}
\mbox{.}
\end{equation}
(the quantity $\tilde{p}$ defined in~\eqref{eq:DEFMOMSCHL} is not multi-valued; this
notation is used to indicate that this quantity has been defined with the help of a multi-valued
$\tilde{S}$). As discussed in more detail in section~\ref{sec:gaugecoupling} this
new 'degree of freedom' is intimately related to the existence of gauge fields.
In contrast to $\tilde{S}$, the second dynamic variable $\rho$ is a physical observable
(in the statistical sense) and is treated as a single-valued function.

The necessary and sufficient condition for single-valuedness of a function
$\tilde{S}(x,t)$ (in a subspace $\mathcal{G} \subseteq \mathcal{R}^{4}$) is
that all second order derivatives of $\tilde{S}(x,t)$ with respect to
$x_{k}$ and $t$ commute with each other (in $\mathcal{G}$)
[see e.g.~\cite{kaempfer:concepts}]. As a consequence, the order of two derivatives
of $\tilde{S}$ with respect to anyone of the variables $x_k,t$  must not be changed.
We introduce the (single-valued) quantities
\begin{equation}
  \label{eq:INC35HEOS}
\tilde{S}_{[j,k]} =
\left[\frac{\partial^{2}\tilde{S}}{\partial x_{j}\partial x_{k}} -\frac{\partial^{2}\tilde{S}}{\partial x_{k}\partial x_{j}} \right], \;\;\;
\tilde{S}_{[0,k]} =
\left[\frac{\partial^{2}\tilde{S}}{\partial t \partial x_{k}} -\frac{\partial^{2}\tilde{S}}{\partial x_{k}\partial t } \right]
\mbox{}
\end{equation}
in order to describe the non-commuting derivatives in the following calculations.

The second of the assumptions listed above has been referred to in I
as 'statistical
conditions'. For the present three-dimensional theory these are obtained
in the same way as in I,  by replacing the observables $x_k(t),\,p_k(t)$ and the force
field $F_k(x(t),\,p(t),\,t)$  of the type 1 theory~(\ref{eq:TVONLIUSCHL}) by averages
$\overline{x_k},\,\overline{p_k}$ and $\overline{F_k}$. This leads to the relations
\begin{gather}
\frac{\mathrm{d}}{\mathrm{d}t} \overline{x_k}  =  \frac{\overline{p_k}}{m}
\label{eq:FIRSTAETSCHL}\\
\frac{\mathrm{d}}{\mathrm{d}t} \overline{p_k}  =
\overline{F_k(x,\,p,\,t)}
\label{eq:SECONAETSCHL}
\mbox{,}
\end{gather}
where the averages are given by the following integrals over the random
variables $x_k,\,p_k$ (which should be clearly distinguished from the type I
observables $x_k(t),\,p_k(t)$ which will not be used any more):
\begin{gather}
\overline{x_k} =  \int_{-\infty}^{\infty} \mathrm{d^{3}} x\, \rho(x,t)\, x_k
\label{eq:ERWAETXKSCHL}\\
\overline{p_k}  =  \int_{-\infty}^{\infty} \mathrm{d^{3}} p\, w(p,t)\, p_k
\label{eq:ERWAETPKSCHL}\\
\overline{F_k(x,\,p,\,t)}  =
  \int_{-\infty}^{\infty} \mathrm{d^{3}} x \, \mathrm{d^{3}} p
\,W(x,\,p,\,t)F_k(x,\,p,\,t)
\label{eq:ERWAETFKXPSCHL}
\mbox{.}
\end{gather}
The time-dependent probability densities $W,\,\rho,\,w$
should be positive semidefinite and normalized to unity, i.e. they
should fulfill the conditions
\begin{equation}
  \label{eq:NCFOPDSSCHL}
 \int_{-\infty}^{\infty} \mathrm{d^{3}} x\, \rho(x,t)= \int_{-\infty}^{\infty} \mathrm{d^{3}} p\, w(p,t)=
 \int_{-\infty}^{\infty} \mathrm{d^{3}} x \, \mathrm{d^{3}} p
\,W(x,\,p,\,t)=1
\mbox{}
\end{equation}
The densities $\rho$ and $w$ may be derived from the fundamental probability
density $W$ by means of the relations
\begin{equation}
  \label{eq:IDTHRAWFBWSCHL}
\rho(x,t)=  \int_{-\infty}^{\infty} \mathrm{d^{3}} p\, W(x,\,p,\,t);\hspace{1cm}
w(p,t)= \int_{-\infty}^{\infty} \mathrm{d^{3}} x \,W(x,\,p,\,t)
\mbox{.}
\end{equation}
The present construction of the statistical conditions~(\ref{eq:FIRSTAETSCHL})
and~(\ref{eq:SECONAETSCHL}) from the type 1 theory~(\ref{eq:TVONLIUSCHL}) shows
two differences as compared to the treatment in I. The first is that we allow now for
a $p-$dependent external force. This leads to a more complicated probability density
$W(x,\,p,\,t)$ as compared to the two decoupled densities $\rho(x,t)$ and $ w(p,t)$ of I.
The second difference, which is in fact related to the first, is the use of a
multi-valued $\tilde{S}(x,t)$.

Note, that the $p-$dependent probability densities $w(p,t)$ and $W(x,\,p,\,t)$ have been
introduced in the above relations in a purely formal way. We defined an expectation
value $\overline{p_k}$ [via Eq.~\eqref{eq:FIRSTAETSCHL}] and assumed [in
Eq.~\eqref{eq:ERWAETPKSCHL} ] that a random variable $p_k$ and a corresponding
probability density $w(p,t)$ exist. But the validity of this assumption is not guaranteed .
There is no compelling conceptual basis for the existence of these quantities in a pure
configuration-space theory. If they exist, they must be defined with the help of additional
considerations (see section 6 of I).  The deeper reason for this problem is that the concept
of measurement of momentum (which is proportional to the time derivative of position) is
ill-defined in a theory whose observables are defined in terms of a large number of
experiments at \emph{one and the same} instant of time (measurement of a derivative
requires measurements at different times). Fortunately, these considerations, which have
been discussed in more detail in I, play not a prominent role [apart from the choice of
$W(x,\,p,\,t)$ discussed in section~\ref{sec:stat-constr-macr}], for the derivation
of Schr\"odinger's equation reported in the present paper\footnote{These considerations
seem relevant for attempts to define phase-space densities, e.g. of the Wigner type, in QT}.

Using the continuity equation~\eqref{eq:CONT3DSCHL} and the statistical
conditions~\eqref{eq:FIRSTAETSCHL} and \eqref{eq:SECONAETSCHL} the
present generalization of the integral equation Eq.~(24) of I may be derived.
The steps leading to this result are very similar to the corresponding steps in I and
may be skipped. The essential difference to the one-dimensional treatment is - apart
from the number of space dimensions - the non-commutativity of the second order
derivatives of $\tilde{S}(x,t)$ leading to non-vanishing quantities
$\tilde{S}_{[j,k]},\,\tilde{S}_{[0,k]}$ defined in Eq.~\eqref{eq:INC35HEOS}.
The result takes the form
\begin{equation}
  \label{eq:NID2S3TBSF}
\begin{split}
-&\int_{-\infty}^{\infty} \mathrm{d}^{3} x
\frac{\partial \rho}{\partial x_{k}}
\left[
\frac{\partial \tilde{S}}{\partial t}
+\frac{1}{2m}
\sum_{j}\left( \frac{\partial \tilde{S}}{\partial x_{j}} \right)^{2}
+V
\right]\\
+&\int_{-\infty}^{\infty} \mathrm{d}^{3} x
\rho
\left[
\frac{1}{m}
\frac{\partial \tilde{S}}{\partial x_{j}}
\tilde{S}_{[j,k]}+\tilde{S}_{[0,k]}
\right] = \overline{F^{(e)}_{k}(x,\,p,\,t)}
\end{split}
\mbox{,}
\end{equation}
In the course of the calculation leading to~\eqref{eq:NID2S3TBSF} it has been assumed that
the macroscopic force $F_k(x,\,p,\,t)$ entering the second statistical
condition~(\ref{eq:SECONAETSCHL}) may be written as a sum of two contributions,
$F_k^{(m)}(x,t)$ and $F^{(e)}_k(x,\,p,\,t)$,
\begin{equation}
  \label{eq:AZ12EKI2T}
F_k(x,\,p,\,t)=F_k^{(m)}(x,t)+F^{(e)}_k(x,\,p,\,t)
\mbox{,}
\end{equation}
where $F_k^{(m)}(x,t)$ takes the form of a negative gradient of a
scalar function $V(x,t)$ (mechanical potential) and $F^{(e)}_k(x,\,p,\,t)$
is the remaining $p-$dependent part.

Comparing Eq.~(\ref{eq:NID2S3TBSF}) with the corresponding formula
obtained in I [see Eq.~(24) of I] we see that two new terms appear
now, the expectation value of the $p-$dependent force on the r.h.s.,
and the second term on the l.h.s. of Eq.~(\ref{eq:NID2S3TBSF}). The
latter is a direct consequence of our assumption of a multi-valued variable
$\tilde{S}$. In section~\ref{sec:stat-constr-macr} it will be shown that
for vanishing multi-valuedness Eq.~(\ref{eq:NID2S3TBSF}) has to agree
with the three-dimensional generalization of the corresponding result
[Eq.~(24) of I] obtained in I. This means that the $p-$dependent term
on the r.h.s. has to vanish too in this limit and indicates a relation
between  multi-valuedness of $\tilde{S}$ and $p-$dependence of
the external force.

\section{Gauge coupling as a consequence of a multi-valued phase}
\label{sec:gaugecoupling}

In this section we study the consequences of the
multi-valuedness [\cite{london:gauge},~\cite{weyl:elektron},~\cite{dirac:quantised}] of the quantity $\tilde{S}(x,t)$ in the present theory. We assume
that $\tilde{S}(x,t)$ may be written as a sum of a single-valued part $S(x,t)$
and a multi-valued part $\tilde{N}$. Then, given that~\eqref{eq:TGT7RMT}
holds, the derivatives of $\tilde{S}(x,t)$ may be written in the form
\begin{equation}
  \label{eq:DOS4MBWIF}
\frac{\partial \tilde{S}}{\partial t}=
\frac{\partial S}{\partial t}+e \Phi,\;\;\;\;\;
\frac{\partial \tilde{S}}{\partial x_{k}}=
\frac{\partial S}{\partial x_{k}}-\frac{e}{c}A_{k}
\mbox{,}
\end{equation}
where the four functions $\Phi$ and $A_{k}$ are proportional to the
derivatives of $\tilde{N}$ with respect to $t$ and $x_{k}$ respectively
(Note the change in sign of $\Phi$ and $A_{k}$ in comparison
to~\cite{klein:schroedingers}; this is due to the fact that the multi-valued
phase is now denoted by $\tilde{S}$). The physical motivations for
introducing the pre-factors $e$ and $c$ in Eq.~(\ref{eq:DOS4MBWIF})
have been extensively discussed
elsewhere, see \cite{kaempfer:concepts},~\cite{klein:schroedingers},
in an electrodynamical context.  In agreement with
Eq.~\eqref{eq:DOS4MBWIF}, $\tilde{S}$ may be written
in the form [\cite{kaempfer:concepts},~\cite{klein:schroedingers}]
\begin{equation}
  \label{eq:AWIZ7MMVF}
\tilde{S}(x,t;\mathcal{C})=S(x,t)-
\frac{e}{c}
\int_{x_{0},t_{0};\mathcal{C}}^{x,t}
\left[\mathrm{d}x_{k}' A_k(x',t')-c\mathrm{d}t' \phi(x',t')\right]
\mbox{,}
\end{equation}
as a path-integral performed along an arbitrary path $\mathcal{C}$
in four-dimensional space; the multi-valuedness of $\tilde{S}$
simply means that it depends not only on $x,t$ but also on the path
$\mathcal{C}$ connecting the points $x_0,t_0$ and $x,t$.

The quantity $\tilde{S}$ cannot be a physical observable because
of its multi-valuedness. The fundamental physical quantities to be
determined by our (future) theory are the four derivatives of
$\tilde{S}$ which will be rewritten here as two observable
fields $-\tilde{E}(x,\,t)$,  $\tilde{p}_k(x,\,t)$,
\begin{gather}
-\tilde{E}(x,\,t)=\frac{\partial S(x,t)}{\partial t}+e \Phi(x,t),
\label{eq:DOJUE8W11}\\
\tilde{p}_k(x,\,t)=\frac{\partial S(x,t)}{\partial x_{k}}-\frac{e}{c}A_{k}(x,t)
 \label{eq:DOJUE8W22}
\mbox{,}
\end{gather}
with dimensions of  energy and momentum respectively.

We encounter a somewhat unusual situation in
Eqs.~(\ref{eq:DOJUE8W11}),~(\ref{eq:DOJUE8W22}): On the one hand
the left hand sides are observables of our theory, on the other hand we
cannot solve our (future) differential equations for these quantities
because of the peculiar multi-valued structure of $\tilde{S}$. We
have to use instead the decompositions as given by the right hand
sides of~(\ref{eq:DOJUE8W11}) and~(\ref{eq:DOJUE8W22}). The latter eight
terms (the four derivatives of $S$ and the four scalar functions  $\Phi$ and $A_{k}$)
are single-valued (in the mathematical sense) but need not be  unique because
only the left hand sides are uniquely determined by the physical situation.
We tentatively assume that the fields $\Phi$ and $A_{k}$ are 'given' quantities in the
sense that they represent an external influence (of 'external forces')
on the considered statistical situation. An actual calculation has
to be performed in such a way that fixed fields $\Phi$ and $A_{k}$ are
chosen and then the differential equations are solved for $S$ (and
$\rho$). However, as mentioned already, what is actually uniquely
determined by the physical situation is the \emph{sum} of the two
terms on the right hand sides of~(\ref{eq:DOJUE8W11})
and~(\ref{eq:DOJUE8W22}). Consequently, a different set of fixed fields
$\Phi^{'}$ and $A_{k}^{'}$ may lead to a physically equivalent, but
mathematically different, solution $S^{'}$ in such a way that the sum of
the new terms [on the right hand sides of~(\ref{eq:DOJUE8W11})
and~(\ref{eq:DOJUE8W22})] is the \emph{same} as the sum of the
old terms. We assume here, that the formalism restores the values of
the physically relevant terms. This implies that the relation between
the old and new terms is given by
\begin{gather}
S^{'}(x,t)=S(x,t)+\varphi(x,t) \label{eq:EEFFRRW11} \\
\Phi^{'}(x,t)=\Phi(x,t)-\frac{1}{e}\frac{\partial \varphi(x,t)}{\partial t}
\label{eq:EEFFRRW22} \\
A_{k}^{'}(x,t)=A_{k}(x,t)+\frac{c}{e}\frac{\partial \varphi(x,t)}{\partial x_k}
\label{eq:EEFFRRW33}
\mbox{,}
\end{gather}
where $\varphi(x,t)$ is an arbitrary, single-valued function of
$x_k,\,t$. Consequently, all 'theories' (differential equations
for $S$ and $\rho$  defined by the assumptions listed in
section~\ref{sec:calcwpt}) should be form-invariant under the
transformations~(\ref{eq:EEFFRRW11})-(\ref{eq:EEFFRRW33}).
These invariance transformations, predicted here from general considerations,
are (using an arbitrary function $\chi=c\varphi/e$ instead of $\varphi$) denoted as
'gauge transformations of the second kind'.

The fields $\Phi(x,t)$ and $A_{k}(x,t)$ describe an external
influence but their numerical value is undefined;  their value at
$x,\,t$ may be changed  according to~(\ref{eq:EEFFRRW22})
and~(\ref{eq:EEFFRRW33}) without changing their physical
effect. Thus, these fields \emph{cannot play a local role} in space
and time like forces and fields in classical mechanics and
electrodynamics. What, then, is the physical meaning of these
fields ? An explanation which seems obvious in the present context
is the following: They describe the
\emph{statistical effect} of an external influence on the
considered system (ensemble of identically prepared
individual particles). The statistical effect of a force field on
an ensemble may obviously \emph{differ} from the local effect of the
same force field on individual particles; thus the very existence
of fields $\Phi$ and $A_{k}$ different from $\vec{E}$ and $\vec{B}$
is no surprise. The second common problem with the interpretation
of the 'potentials' $\Phi$ and $A_{k}$ is their non-uniqueness. It is hard to
understand that a quantity ruling the behavior of individual particles
should not be uniquely defined. In contrast, this non-uniqueness is much
easier to accept if $\Phi$ and $A_{k}$ rule the behavior of
ensembles instead of individual particles. We have no problem to
accept the fact that a function that represents a global (integral)
effect may have many different local realizations.

It seems that this interpretation of the potentials $\Phi$ and
$A_{k}$ is highly relevant for the interpretation of the
effect found by \cite{aharonov.bohm:significance}.
If QT is interpreted as a theory about individual particles, the Aharonov-Bohm
effects imply that a charged particle may be influenced in a nonlocal
way by electromagnetic fields in inaccessible regions. This paradoxical
prediction, which is however in strict agreement with QT, led even to
a discussion about the reality of these
effects, see~\cite{bocchieri.loinger:nonexistence},~\cite{roy:condition},~\cite{klein:nonexistence},~\cite{peshkin.tonomura:aharonov}.
A statistical interpretation of the potentials has apparently never been suggested,
neither in the vast literature about the Aharonov-Bohm effect nor in papers promoting
the statistical interpretation of QT; most physicists discuss this nonlocal
'paradox' from the point of view of 'the wave function of a single electron'.
Further comments on this point may be found in section~\ref{sec:discussion}.

The expectation value $\overline{F^{(e)}_{k}(x,\,p,\,t)}$  on
the right hand side of~(\ref{eq:NID2S3TBSF}) is to be calculated
using local, macroscopic forces whose functional form is still unknown.
Both the potentials and these local forces represent an external influence,
and it is reasonable to assume that the (nonlocal) potentials are the statistical
representatives of the local forces on the r.h.s. of Eq.~(\ref{eq:NID2S3TBSF}).
The latter have to be determined by the potentials but must be
uniquely defined at each space-time point. The gauge-invariant fields
\begin{equation}
  \label{eq:DFS32WUSS}
E_{k}=-\frac{1}{c}\frac{\partial A_{k}}{\partial t}-\frac{\partial\Phi}{\partial x_{k}},\;\;\;\;\;\;
B_{k}= \epsilon_{kij}\frac{\partial A_{j}}{\partial x_{i}}
\mbox{,}
\end{equation}
fulfill these requirements. As a consequence of the defining
relations~(\ref{eq:DFS32WUSS}) they obey automatically the homogeneous
Maxwell equations.

In a next step we rewrite the second term on the l.h.s. of
Eq.~(\ref{eq:NID2S3TBSF}). The commutator terms~(\ref{eq:INC35HEOS})
take the form
\begin{equation}
  \label{eq:DCT88SMBE}
\tilde{S}_{[0,k]} = -e \left(
\frac{1}{c}
\frac{\partial A_{k}}{\partial t}+
\frac{\partial\Phi}{\partial x_{k}} \right),\;\;\;\;
\tilde{S}_{[j,k]} = \frac{e}{c} \left(
\frac{\partial A_{j}}{\partial x_{k}}-
\frac{\partial A_{k}}{\partial x_{j}} \right)
\mbox{.}
\end{equation}
As a consequence, they may be expressed in terms of the local
fields~(\ref{eq:DFS32WUSS}), which have been
introduced above for reasons of gauge-invariance.
Using~(\ref{eq:DCT88SMBE}),~(\ref{eq:DFS32WUSS}) and the
relation~(\ref{eq:DOJUE8W22}) for the momentum field,
Eq.~(\ref{eq:NID2S3TBSF}) takes the form
\begin{equation}
  \label{eq:NIJUR4TBSF}
\begin{split}
-&\int_{-\infty}^{\infty} \mathrm{d}^{3} x
\frac{\partial \rho}{\partial x_{k}}
\left[
\frac{\partial \tilde{S}}{\partial t}
+\frac{1}{2m}
\sum_{j}\left( \frac{\partial \tilde{S}}{\partial x_{j}} \right)^{2}
+V
\right]\\
+&\int_{-\infty}^{\infty} \mathrm{d}^{3} x \rho
\left[\frac{e}{c}  \epsilon_{kij} \tilde{v}_{i} B_{j} + eE_{k}
\right] = \overline{F^{(e)}_{k}(x,\,p,\,t)}
\end{split}
\mbox{,}
\end{equation}
with a velocity field defined by $\tilde{v}_{i}=\tilde{p}_{i}/m$.
Thus, the new terms on the l.h.s. of~(\ref{eq:NIJUR4TBSF}) - stemming from
the multi-valuedness of $\tilde{S}$ - take the form of an expectation value
(with $\mathcal{R}^{3}$ as sample space) of the Lorentz force field
\begin{equation}
  \label{eq:LOZU5FF}
\vec{F}_{L}(x,t)=e\vec{E}(x,t)+\frac{e}{c}\vec{\tilde{v}}(x,t)
\times \vec{B}(x,t)
\mbox{,}
\end{equation}
if the particle velocity is identified with the velocity
field $\vec{\tilde{v}}(x,t)$.

The above steps imply a relation between potentials and local fields.
From the present statistical (nonlocal) point of view the potentials are more
fundamental than the local fields. In contrast, considered from the point of
view of macroscopic physics, the local fields are the physical quantities of
primary importance and the potentials may (or may not) be introduced for
mathematical convenience.

\section{A constraint for forces in statistical theories}
\label{sec:stat-constr-macr}

Let us discuss now the nature of the macroscopic forces
$F_{k}^{(e)}(x,\,p,\,t)$ entering the expectation value on the
r.h.s. of Eq.~(\ref{eq:NIJUR4TBSF}). In our type I parent theory,
classical mechanics, there are no constraints for the possible functional
form of $F_{k}^{(e)}(x,\,p,\,t)$. However, this need not be true in
the present statistical framework. As a matter of
fact, the way the mechanical potential $V(x,t)$ entered the differential
equation for $S$ (in the previous work I) indicates already that such
constraints do actually exist. Let us recall that in I we tacitly restricted
the class of forces to those derivable from a potential $V(x,t)$. If
we eliminate this restriction and admit arbitrary forces, with
components $F_{k}(x,t)$, we obtain instead of the above
relation~(\ref{eq:NIJUR4TBSF}) the simpler relation [Eq.~(24) of I,
generalized to three dimensions and arbitrary forces of the
form $F_{k}(x,t)$]
\begin{equation}
  \label{eq:DZJU623F}
 -\int_{-\infty}^{\infty} \mathrm{d}^{3} x\,\frac{\partial\rho}{\partial x_{k}}\,
\left[
\frac{1}{2m}\sum_{j} \left(\frac{\partial S}{\partial x_{j}} \right)^{2}+
\frac{\partial S}{\partial t}
\right]=
\int_{-\infty}^{\infty} \mathrm{d} x \rho F_{k}(x,t)
\mbox{.}
\end{equation}
This is a rather complicated integro-differential equation for our
variables $\rho(x,t)$ and $S(x,t)$. We assume now, using mathematical
simplicity as a guideline, that Eq.~\eqref{eq:DZJU623F} can be written
in the common form of a local differential equation. This assumption
is of course not evident; in principle the laws of physics could
be integro-differential equations or delay differential equations or
take an even more complicated mathematical form. Nevertheless, this
assumption seems rather weak considering the fact that all fundamental
laws of physics take this 'simple' form. Thus, we postulate that
Eq.~\eqref{eq:DZJU623F} is equivalent to a differential equation
\begin{equation}
  \label{eq:DZ3MSV93F}
\frac{1}{2m}\sum_{j} \left(\frac{\partial S}{\partial x_{j}} \right)^{2}+
\frac{\partial S}{\partial t} + T = 0
\mbox{,}
\end{equation}
where the unknown term $T$ describes the influence of the force
$F_{k}$ but may also contain other contributions. Let us write
\begin{equation}
  \label{eq:JUE23PO2W}
T=-L_{0}+V
\mbox{,}
\end{equation}
where $L_{0}$ does not depend on $F_{k}$, while $V$ depends on it and
vanishes for $F_{k} \to 0$. Inserting~\eqref{eq:DZ3MSV93F} and
\eqref{eq:JUE23PO2W} in~\eqref{eq:DZJU623F} yields
\begin{equation}
  \label{eq:DTWQ84O7F}
\int \mathrm{d}^{3}x \,\frac{\partial\rho}{\partial x_{k}}\,
\left( -L_{0}+V \right)= \int \mathrm{d}^{3}x\, \rho F_{k}(x,t)
\mbox{.}
\end{equation}
For $F_{k} \to 0$ Eq.~\eqref{eq:DTWQ84O7F} leads to the relation
\begin{equation}
  \label{eq:DQWEI977F}
\int \mathrm{d}^{3}x \,\frac{\partial\rho}{\partial x_{k}}\,L_{0}= 0
\mbox{,}
\end{equation}
which remains true for finite forces because $L_{0}$ does not
depend on $F_{k}$. Finally, performing a partial integration, we see
that a relation
\begin{equation}
  \label{eq:ESZURVF5ZW}
F_{k}=-\frac{\partial V}{\partial x_{k}}+s_{k},\; \; \;
\int_{-\infty}^{\infty} \mathrm{d}^{3}x\,\rho s_{k}=0
\mbox{,}
\end{equation}
exists between  $F_{k}$ and $V$, with a vanishing expectation
value of the (statistically irrelevant) functions $s_{k}$. This
example shows that the restriction to gradient fields, made above and
in I, is actually not necessary. We may \emph{admit} force fields which
are arbitrary functions of $x$ and $t$; the statistical conditions
(which play now the role of a 'statistical constraint') eliminate
automatically all forces that cannot be written after statistical
averaging as gradient fields.

This is very interesting and indicates the possibility that the
present statistical assumptions leading to Schr\"odinger's equation
may also be responsible, at least partly, for the structure of the
real existing (gauge) interactions of nature.

Does this statistical constraint also work in the present $p-$dependent
case ? We assume that the force in~(\ref{eq:NIJUR4TBSF}) is a standard
random variable with the configuration space as sample space (see the
discussion in section 4 of I) and that the variable $p$
in $F_{k}^{(e)}(x,\,p,\,t)$ may consequently be replaced by the
field  $\tilde{p}(x,t)$ [see~(\ref{eq:DOJUE8W22})]. Then, the
expectation value on the r.h.s. of~(\ref{eq:NIJUR4TBSF}) takes the form
\begin{equation}
  \label{eq:DEHDF5DIMW}
\overline{F^{(e)}_{k}(x,\,p,\,t)}=
\int_{-\infty}^{\infty} \mathrm{d}^{3} x \rho(x,t)
H_{k}(x,\frac{\partial \tilde{S}(x,t)}{\partial x},t)
\mbox{.}
\end{equation}
The second term on the l.h.s. of~(\ref{eq:NIJUR4TBSF}) has the \emph{same}
form. Therefore, the latter may be eliminated by writing
\begin{equation}
  \label{eq:WWH45HDSTL}
H_{k}(x,\frac{\partial \tilde{S}}{\partial x},t )=
\frac{e}{c}  \epsilon_{kij}
\frac{1}{m} \frac{\partial \tilde{S}}{\partial x_{i}}
B_{j} + eE_{k}
+h_{k}(x,\frac{\partial \tilde{S}}{\partial x},t )
\mbox{,}
\end{equation}
with $h_{k}(x,p,t)$ as our new unknown functions. They obey the
simpler relations
\begin{equation}
  \label{eq:NISSD4BSF}
-\int_{-\infty}^{\infty} \mathrm{d}^{3} x
\frac{\partial \rho}{\partial x_{k}}
\left[
\frac{\partial \tilde{S}}{\partial t}
+\frac{1}{2m}
\sum_{j}\left( \frac{\partial \tilde{S}}{\partial x_{j}} \right)^{2}
+V
\right] =
\int_{-\infty}^{\infty} \mathrm{d}^{3} x \rho
h_{k}(x,\frac{\partial \tilde{S}}{\partial x},t)
\mbox{.}
\end{equation}
On a first look this condition for the allowed forces looks similar to the
$p-$independent case [see~(\ref{eq:DZJU623F})]. But the dependence of
$h_{k}$ on $x,t$ cannot be considered as 'given' (externally controlled),
as in the $p-$independent case, because it contains now the unknown
$x,t$-dependence of the derivatives of $\tilde{S}$. We may
nevertheless try to incorporate the r.h.s by adding a term
$\tilde{T}$ to the bracket which depends on the derivatives of
the multivalued quantity $\tilde{S}$. This leads to the condition
\begin{equation}
  \label{eq:ES2ZODF222}
h_{k}(x,\frac{\partial \tilde{S}}{\partial x},t)=
-\frac{\partial \tilde{T}(x,\frac{\partial \tilde{S}}{\partial x},t)}
{\partial x_{k}}+s_{k},\; \; \;
\int_{-\infty}^{\infty} \mathrm{d}^{3}x\,\rho s_{k}=0
\mbox{.}
\end{equation}
But this relation cannot be fulfilled for nontrivial
$h_{k},\,\tilde{T}$ because the derivatives of $\tilde{S}$ cannot
be subject to further constraints beyond those given by the
differential equation; on top of that the derivatives with
regard to $x$ on the r.h.s. create higher order derivatives
of $\tilde{S}$ which are not present at the l.h.s. of
Eq.~(\ref{eq:ES2ZODF222}). The only possibility to
fulfill this relation is for constant
$\frac{\partial \tilde{S}}{\partial x}$, a special case which
has in fact already be taken into account by adding the mechanical
potential $V$. We conclude that the statistical constraint
leads to $h_{k}=\tilde{T}=0$ and that the statistical
condition~(\ref{eq:NISSD4BSF}) takes the form
\begin{equation}
  \label{eq:NIFF9OTFBSF}
-\int \mathrm{d}^{3} x
\frac{\partial \rho}{\partial x_{k}}
\left[
\frac{\partial \tilde{S}}{\partial t}
+\frac{1}{2m}
\sum_{j}\left( \frac{\partial \tilde{S}}{\partial x_{j}} \right)^{2}
+V
\right] =0
\mbox{.}
\end{equation}

Thus, only a mechanical potential and  the four electrodynamic
potentials are compatible with the statistical constraint and
will consequently - assuming that the present statistical
approach reflects a fundamental principle of nature - be realized in
nature. As is well known all existing interactions follow (sometimes
in a generalized form) the gauge coupling scheme derived above.
The statistical conditions imply not only Schr\"odinger's equation
but also the form of the (gauge) coupling to external influences
and the form of the corresponding local force, the Lorentz force,
\begin{equation}
  \label{eq:LOFFUNW}
\vec{F}_{L}=e\vec{E}+\frac{e}{c}\vec{v} \times \vec{B}
\mbox{,}
\end{equation}
if the particle velocity $\vec{v}$  is identified with the velocity field
$\vec{\tilde{v}}(x,t)$.

In the present derivation the usual order of proceeding is just
inverted. In the conventional deterministic treatment the form of
the local forces (Lorentz force), as taken from experiment, is used
as a starting point. The potentials are introduced afterwards, in
the course of a transition to a different formal framework (Lagrange
formalism). In the present approach the fundamental assumptions are
the statistical conditions. Then, taking into account an existing
mathematical freedom (multi-valuedness of a variable) leads to the
introduction of potentials. From these, the shape of the macroscopic
(Lorentz) force can be derived, using the validity of the statistical
conditions as a constraint.

\section{Fisher information as the hallmark of quantum theory}
\label{sec:fisher-information}
The remaining nontrivial task is the derivation of a local differential equation
for $S$ and $\rho$ from the integral equation~(\ref{eq:NIFF9OTFBSF}). As our
essential constraint we will use, besides general principles of
simplicity (like homogeneity and isotropy of space) the principle
of maximal disorder, as realized by the requirement of minimal Fisher
information. Using the abbreviation
\begin{equation}
  \label{eq:DG4DGFL3NEU}
\bar{L}(x,t)=\frac{\partial \tilde{S}}{\partial t}+\frac{1}{2m}\left( \frac{\partial \tilde{S}(x,t)}{\partial x}\right)^{2}
+V(x,t)
\mbox{,}
\end{equation}
the general solution of~(\ref{eq:NIFF9OTFBSF}) may be written in the
form
\begin{equation}
  \label{eq:DAL2L9IWF}
\frac{\partial \rho}{\partial x_{k}}
\bar{L}(x,t) = G_{k}(x,t)
\mbox{,}
\end{equation}
where the three functions $G_{k}(x,t)$ have to vanish upon integration
over $\mathcal{R}^{3}$ and are otherwise arbitrary. If we restrict
ourselves to an isotropic law, we may write
\begin{equation}
  \label{eq:WR23OTAIL}
G_{k}(x,t)=\frac{\partial \rho}{\partial x_{k}}
L_{0}
\mbox{.}
\end{equation}
Then, our problem is to find a function $L_{0}$ which
fulfills the differential equation
\begin{equation}
  \label{eq:ZUW78ERHD}
\bar{L}(x,t)-L_{0}=0
\mbox{,}
\end{equation}
and condition~\eqref{eq:DQWEI977F}. The method used in I for a
one-dimensional situation, to determine $L_{0}$ from the requirement
of minimal Fisher information, remains essentially unchanged in the
present three-dimensional case. The reader is referred to the detailed
explanations reported in I.

In I it has been shown that this principle of maximal disorder leads
to an anomalous variational problem and to the following conditions
for our unknown function $L_{0}$:
\begin{gather}
\bar{L}(x,t)-
L_{0}\left(\rho,\frac{\partial \rho}{\partial x},
\frac{\partial^{2} \rho}{\partial x \partial x} \right)
=0  \label{eq:EEFFRRWWW2} \\
\delta \int \mathrm{d}^{3} x\rho
\left[ \bar{L}(x,t)-
L_{0} \left( \rho,\frac{\partial \rho}{\partial x},
\frac{\partial^{2} \rho}{\partial x \partial x} \right)
\right] = 0
\label{eq:EEFFRRWXYZ2}
\mbox{,}
\end{gather}
where $L_{0}$ contains only derivatives of $\rho$ up to second
order and does not explicitely depend on $x,\,t$. If
Eq.~(\ref{eq:EEFFRRWWW2}) is taken into account, the Euler-Lagrange
equations of the variational problem~(\ref{eq:EEFFRRWXYZ2})
lead to the following differential equation
\begin{equation}
  \label{eq:NK17DEDGLV}
-\frac{\partial}{\partial x_{k}}  \frac{\partial}{\partial x_{i}}
\frac{\partial \beta}
{\partial\left(
\frac{\partial^{2} \rho}{\partial x_{k} \partial x_{i}}
\right)}
+\frac{\partial}{\partial x_{k}}
\frac{\partial \beta}
{\partial\left(
\frac{\partial \rho}{\partial x_{k} } \right)}
-\frac{\partial \beta}{\partial \rho}
+\frac{\beta}{\rho}=0
\mbox{}
\end{equation}
for the variable $\beta=\rho L_{0}$. Eq.~(\ref{eq:NK17DEDGLV}) is a
straightforward generalization of the corresponding one-dimensional
relation [equation (68) of I] to three spatial dimensions.

Besides~(\ref{eq:NK17DEDGLV}) a further (consistency) condition
exists, which leads to a simplification of the problem. The function
$L_{0}$ may depend on second order derivatives of $\rho$ but this
dependence must be of a special form not leading to any terms in the
Euler-Lagrange equations [according to~(\ref{eq:EEFFRRWWW2}) our
final differential equation for $S$ and $\rho$ must not contain
higher than second order derivatives of $\rho$]. Consequently,
the first term in Eq.~(\ref{eq:NK17DEDGLV}) (as well as the sum of
the remaining terms) has to vanish separately and~(\ref{eq:NK17DEDGLV})
can be replaced by the two equations
\begin{gather}
 \frac{\partial}{\partial x_{k}}  \frac{\partial}{\partial x_{i}}
\frac{\partial \beta}
{\partial\left(
\frac{\partial^{2} \rho}{\partial x_{k} \partial x_{i}}
\right)}
=0   \label{eq:HHWQHW11} \\
 \frac{\partial}{\partial x_{k}}
\frac{\partial \beta}
{\partial\left(
\frac{\partial \rho}{\partial x_{k} } \right)}
-\frac{\partial \beta}{\partial \rho}
+\frac{\beta}{\rho}=0
\label{eq:EHHWQRW22}
\mbox{.}
\end{gather}
In I a new derivation of Fisher's functional has been obtained, using
the general solution of the one-dimensional version
of~(\ref{eq:NK17DEDGLV}), as well as the so-called composition law.
In the present three-dimensional situation we set ourselves a less
ambitious aim. We know that Fisher's functional describes the maximal
amount of disorder. If we are able to find a solution
of~(\ref{eq:HHWQHW11}),~(\ref{eq:EHHWQRW22}) that agrees with
this functional (besides 'null-terms' giving no contribution to
the Euler-Lagrange equations) then we will accept it as our correct
solution. It is easy to see that this solution is given by
\begin{equation}
  \label{eq:DES3SL5HLL}
L_{0}=B_{0}\left[
-\frac{1}{2\rho^{2}}\sum_j \left(\frac{\partial \rho}
{\partial x_{j}} \right)^{2}
+\frac{1}{\rho}\sum_j \frac{\partial^{2} \rho}
{\partial x_{j}^{2}}
 \right]
\mbox{,}
\end{equation}
where $B_{0}$ is an arbitrary constant. Eq.~(\ref{eq:DES3SL5HLL})
presents again the three-dimensional (and isotropic) generalization
of the one-dimensional result obtained in I. By means of the identity
\begin{equation}
  \label{eq:ZU23A98SI}
\frac{\partial}{\partial x_{i}}
\frac{\partial \sqrt{\rho} }{\partial x_{i}}
\frac{\partial \sqrt{\rho} }{\partial x_{k}}=
\frac{\partial \sqrt{\rho} }{\partial x_{k}}
\frac{\partial}{\partial x_{i}}
\frac{\partial \sqrt{\rho} }{\partial x_{i}}+
\frac{1}{2}
\frac{\partial}{\partial x_{k}}
\frac{\partial \sqrt{\rho} }{\partial x_{i}}
\frac{\partial \sqrt{\rho} }{\partial x_{i}}
\mbox{,}
\end{equation}
it is easily verified that the solution~(\ref{eq:DES3SL5HLL}) obeys
also condition~(\ref{eq:DQWEI977F}). Using the
decomposition~(\ref{eq:DOS4MBWIF}) and renaming $B$ according to
$B=\hbar^{2}/4m$, the continuity equation~(\ref{eq:CONT3DSCHL}) and
the second differential equation~(\ref{eq:EEFFRRWWW2}) respectively,
take the form
\begin{gather}
\frac{\partial \rho}{\partial t}+\frac{\partial}{\partial x_{k}}
\frac{\rho}{m} \left(
\frac{\partial S}{\partial x_{k}}-\frac{e}{c} A_{k}
\right)=0
  \label{eq:CONH3TMF}
\mbox{,} \\
\frac{\partial S}{\partial t}+e\phi + \frac{1}{2m}
\sum_{k}
\left(
\frac{\partial S}{\partial x_{k}}-\frac{e}{c} A_{k}
\right)^{2} +V  -
\frac{\hbar^2}{2m}\frac{\triangle\sqrt{\rho}}{\sqrt{\rho}}
=0
\label{eq:QHJH34MF}
\mbox{.}
\end{gather}
The function $S$ occurring in~(\ref{eq:CONH3TMF}),~(\ref{eq:QHJH34MF}) is
single-valued but not unique (not gauge-invariant). If now the
complex-valued variable
\begin{equation}
  \label{eq:HDIS3JWUI}
\psi = \sqrt{\rho}\mathrm{e}^{\imath \frac{S}{\hbar}}
\mbox{,}
\end{equation}
is introduced, the two equations~(\ref{eq:CONH3TMF}),~(\ref{eq:QHJH34MF})
may be written in compact form as real and imaginary parts of the
linear differential equation
\begin{equation}
  \label{eq:DGMAENE}
 \big(\frac{\hbar}{\imath} \frac{\partial}{\partial t}
+e\phi \big)\psi
+ \frac{1}{2 m }
\big(\frac{\hbar}{\imath} \frac{\partial}{\partial \vec{x}} -
 \frac{e}{c} \vec{A} \big)^{2}\psi
+V\psi=0
\mbox{,}
\end{equation}
which completes our derivation of Schr\"odinger's equation in the presence
of a gauge field.

Eq.~(\ref{eq:DGMAENE}) is in manifest gauge-invariant
form. The gauge-invariant derivatives of $\tilde{S}$  with respect
to $t$ and $\vec{x}$ correspond to the two brackets in~(\ref{eq:DGMAENE}).
In particular, the canonical momentum  $\partial S/ \partial \vec{x}$
corresponds to the momentum operator proportional to
$\partial / \partial \vec{x}$. Very frequently, Eq.~(\ref{eq:DGMAENE})
is written in the form
\begin{equation}
  \label{eq:SCH96RG2IFU}
-\frac{\hbar}{\imath} \frac{\partial}{\partial t} \psi=
H \psi,
\mbox{,}
\end{equation}
with the Hamilton operator
\begin{equation}
  \label{eq:HA3UZM9OPR}
H =
\frac{1}{2 m }
\big(\frac{\hbar}{\imath} \frac{\partial}{\partial \vec{x}} -
 \frac{e}{c} \vec{A} \big)^{2}
+V +e\phi
\mbox{, }
\end{equation}

Our final result, Eqs.~\eqref{eq:SCH96RG2IFU},~\eqref{eq:HA3UZM9OPR}, agrees
with the result of the conventional quantization procedure. In its simplest form, the latter
starts from the classical relation $H(x,p)=E$, where $H(x,p)$ is the Hamiltonian of
a classical particle in a conservative force field, and  $E$ is its energy. To perform a
"canonical quantization" means to replace $p$ and $E$ by differential expressions according
to~\eqref{eq:DTR43QUPR} and let then act both sides of the equation $H(x,p)=E$ on
states $\psi$ of a function space. The 'black magic' involved in this process has been
eliminated, or at least dramatically reduced, in the present approach, where
Eqs.~\eqref{eq:SCH96RG2IFU},~\eqref{eq:HA3UZM9OPR} have been derived
from a set of assumptions which can all be interpreted in physical terms.

The Hamiltonian~\eqref{eq:HA3UZM9OPR} depends on the potentials $\Phi$ and
$\vec{A}$ and is consequently a non-unique (not gauge-invariant) mathematical
object. The same is true for the time-development operator $U(H)$ which is an
operator function of $H$, see e.g.~\cite{kobe:time-evolution}. This non-uniqueness
is a problem if $U(H)$ is interpreted as a quantity ruling the time-evolution of
a single particle. It is no problem from the point of view of the SI where
$H$ and $U(H)$ are primarily convenient mathematical objects which occur
in a natural way if the time-dependence of statistically relevant (uniquely defined)
quantities, like expectation values and transition probabilities, is to be calculated.

\section{Spin as a statistical degree of freedom}
\label{sec:spin}

Spin is generally believed to be a phenomenon of quantum-theoretic origin. For a
long period of time, following Dirac's derivation of his relativistic equation, it was
also believed to be essentially of relativistic origin. This has changed since the work
of~\cite{schiller:spinning},~\cite{levy-leblond:nonrelativistic},~\cite{arunsalam:hamiltonians},~\cite{gould:intrinsic},~\cite{reginatto:pauli} and others, who showed that spin may be derived
entirely in the framework of non-relativistic QT without using any relativistic concepts.
Thus, a new derivation of non-relativistic QT like the present one should also include
a derivation of the phenomenon of spin. This will be done in this and the next two
sections.

A simple idea to extend the present theory is to assume that sometimes -
under certain external conditions to be identified later - a situation
occurs where the behavior of our statistical ensemble of particles cannot
longer be described by $\rho,\,S$ alone but requires, e.g., the double
number of field variables; let us denote these by
$\rho_1,\,S_1,\,\rho_2,\,S_2$ (we restrict ourselves here to spin one-half).
The relations defining this generalized theory should be formulated
in such a way that the previous relations are obtained in the appropriate limits.
One could say that we undertake an attempt to introduce a new (discrete)
degree of freedom for the ensemble. If we are able to derive a non-trivial
set of differential equations - with coupling between $\rho_1,\,S_1$ and
$ \rho_2,\,S_2$ - then such a degree of freedom could exist in nature.

Using these guidelines, the basic equations of the generalized theory can be
easily formulated. The probability density and probability current take the form
$\rho = \rho_1 + \rho_2$ and $\vec{j}=\vec{j}_1+\vec{j}_2$, with
$\vec{j}_i\;$($i=1,2$) defined in terms of $\rho_i,\,S_i$ exactly as before
(see section~\ref{sec:calcwpt}). Then, the continuity equation is given by
\begin{equation}
  \label{eq:contspin}
\frac{\partial (\rho_1+\rho_2)}{\partial t}+
\frac{\partial}{\partial x_l}
\left(
\frac{\rho_1}{m}\frac{\partial \tilde{S}_1}{\partial x_l}
+\frac{\rho_2}{m}\frac{\partial \tilde{S}_2}{\partial x_l}
\right)=0
\mbox{,}
\end{equation}
where we took the possibility of multi-valuedness of the ``phases``
already into account, as indicated by the notation $\tilde{S}_i$.
The statistical conditions are given by the two relations
\begin{gather}
\frac{\mathrm{d}}{\mathrm{d}t} \overline{x_k}  =  \frac{\overline{p_k}}{m}
\label{eq:FIRSTAETSPIN}\\
\frac{\mathrm{d}}{\mathrm{d}t} \overline{p_k}  =
\overline{F_k^{(T)}(x,\,p,\,t)}
\label{eq:SECONAETSPIN}
\mbox{,}
\end{gather}
which are similar to the relations used previously (in
section~\ref{sec:calcwpt} and in I), and by an additional equation
\begin{equation}
  \label{eq:ANA35EQUT}
\frac{\mathrm{d}}{\mathrm{d}t} \overline{s_k}  =
\overline{F_k^{(R)}(x,\,p,\,t)}
\mbox{,}
\end{equation}
which is required as a consequence of our larger number of dynamic
variables. Eq.~(\ref{eq:ANA35EQUT}) is best explained later; it is
written down here for completeness. The forces $F_k^{(T)}(x,\,p,\,t)$
and $F_k^{(R)}(x,\,p,\,t)$ on the r.h.s. of~(\ref{eq:SECONAETSPIN})
and~(\ref{eq:ANA35EQUT}) are again subject to the ``statistical
constraint``, which has been defined in section~\ref{sec:gaugecoupling}.
The expectation values are defined as
in~(\ref{eq:ERWAETXKSCHL})-(\ref{eq:ERWAETFKXPSCHL}).

Performing mathematical manipulations similar to the
ones reported in section~\ref{sec:calcwpt},
the l.h.s. of Eq.~(\ref{eq:SECONAETSPIN}) takes the form
\begin{equation}
  \label{eq:AHK1LUA9IB}
\begin{split}
\frac{\mathrm{d}}{\mathrm{d}t} \overline{p_k}  &=
\int \mathrm{d}^{3} x
\Big[
\frac{\partial \rho_1}{\partial t}
\frac{\partial \tilde{S}_1}{\partial x_{k}}
+\frac{\partial \rho_2}{\partial t}
\frac{\partial \tilde{S}_2}{\partial x_{k}}\\
& - \frac{\partial \rho_1}{\partial x_{k}}
\frac{\partial \tilde{S}_1}{\partial t}
- \frac{\partial \rho_2}{\partial x_{k}}
\frac{\partial \tilde{S}_2}{\partial t}
+\rho_1 \tilde{S}_{[0,k]}^{(1)}
+\rho_2 \tilde{S}_{[0,k]}^{(2)}
\Big] \mbox{,}
\end{split}
\end{equation}
where the quantities $ \tilde{S}_{[j,k]}^{(i)},\,i=1,2$ are defined as
above [see Eq.~(\ref{eq:INC35HEOS})] but with  $\tilde{S}$ replaced by
$\tilde{S}_i$.

Let us write now $\tilde{S}$ in analogy to
section~\ref{sec:calcwpt} in the form $\tilde{S}_i=S_i+\tilde{N}_i$,
as a sum of a single-valued part $S_i$ and a multi-valued part
$\tilde{N}_i$. If $\tilde{N}_1$ and $\tilde{N}_2$ are to represent
an external influence, they must be identical and a single
multi-valued part $\tilde{N}=\tilde{N}_1=\tilde{N}_2$ may be used
instead. The derivatives of $\tilde{N}$ with respect to $t$ and
$x_k$ must be single-valued and we may write
\begin{equation}
  \label{eq:DHI9DD8HIN}
\frac{\partial \tilde{S}_i}{\partial t}=
\frac{\partial S_i}{\partial t}+e \Phi,\;\;\;\;\;
\frac{\partial \tilde{S}_i}{\partial x_{k}}=
\frac{\partial S_i}{\partial x_{k}}-\frac{e}{c}A_{k}
\mbox{,}
\end{equation}
using the same familiar electrodynamic notation as in
section~\ref{sec:calcwpt}. In this way we arrive at
eight single-valued functions to describe the external
conditions and the dynamical state of our system,
namely $\Phi,\,A_{k}$ and $\rho_i,\,S_i$.

In a next step we replace $\rho_i,\,S_i$ by new dynamic variables
$\rho,\,S,\,\vartheta,\,\varphi$ defined by
\begin{equation}
  \label{eq:DS12TRF8FU}
\begin{split}
&\rho_1=\rho \cos^{2}\frac{\vartheta}{2},
\hspace{1.6cm}S_1=S+\frac{\hbar}{2}\varphi, \\
&\rho_2=\rho \sin^{2}\frac{\vartheta}{2},
\hspace{1.6cm}S_2=S-\frac{\hbar}{2}\varphi.
\end{split}
\end{equation}
A transformation similar to Eq.~(\ref{eq:DS12TRF8FU}) has
been introduced by~\cite{takabayasi:vector} in his
reformulation of Pauli's equation. Obviously, the variables
$S,\,\rho$ describe 'center of mass' properties (which are common to
both states $1$ and $2$) while $\vartheta,\,\varphi$ describe
relative (internal) properties of the system.

The dynamical variables $S,\,\rho$ and $\vartheta,\,\varphi$ are
not decoupled from each other. It turns out (see below) that
the influence of $\vartheta,\,\varphi$ on $S,\,\rho$ can be described
in a (formally) similar way as the influence of an external
electromagnetic field if a 'vector potential' $\vec{A}^{(s)}$ and a
'scalar potential' $\phi^{(s)}$, defined by
\begin{equation}
  \label{eq:HIDV32PESSF}
A_{l}^{(s)}= -\frac{\hbar c}{2e} \cos\vartheta \frac{\partial
\varphi}{\partial x_l},\hspace{1cm}
\phi^{(s)} = \frac{\hbar}{2e} \cos\vartheta \frac{\partial
\varphi}{\partial t}
\mbox{,}
\end{equation}
are introduced. Denoting these fields as 'potentials', we should bear
in mind that they are not externally controlled but defined in terms of
the internal dynamical variables. Using the abbreviations
\begin{equation}
  \label{eq:ZUWA94BRR}
\hat{A_l}=A_l+A_{l}^{(s)},\hspace{1cm}
\hat{\phi}=\phi+\phi^{(s)}
\mbox{,}
\end{equation}
the second statistical condition~(\ref{eq:SECONAETSPIN}) can be written
in the following compact form
\begin{equation}
  \label{eq:H7TRJS4DZ}
\begin{split}
&-\int \mathrm{d}^{3} x
\frac{\partial \rho}{\partial x_{l}}
\bigg[
\bigg( \frac{\partial S}{\partial t}+e \hat{\phi} \bigg)
+\frac{1}{2m}
\sum_{j}
\bigg(
\frac{\partial S}{\partial x_{j}} - \frac{e}{c} \hat{A_j}
\bigg)^{2}
\bigg]\\
&+\int \mathrm{d}^{3} x
\rho
\bigg[
-\frac{e}{c} v_j
\bigg(
 \frac{\partial \hat{A}_l}{\partial x_{j}}
-\frac{\partial \hat{A}_j}{\partial x_{l}}
\bigg)
 -\frac{e}{c} \frac{\partial \hat{A}_l}{\partial t}
 -e \frac{\partial \hat{\phi}}{\partial x_{l}}
\bigg] \\
&= \overline{F^{(T)}_{l}(x,\,p,\,t)}
=\int \mathrm{d}^{3} x \rho F^{(T)}_{l}(x,\,p,\,t)
\mbox{,}
\end{split}
\end{equation}
which shows a formal similarity to the spinless case
[see~(\ref{eq:NID2S3TBSF}) and~(\ref{eq:DCT88SMBE})].
The components of the velocity field in~(\ref{eq:H7TRJS4DZ})
are given by
\begin{equation}
  \label{eq:DSS3DKDXVF}
v_j=\frac{1}{m} \bigg(
\frac{\partial S}{\partial x_{j}} - \frac{e}{c} \hat{A_j}
\bigg)
\mbox{.}
\end{equation}
If now fields $E_l,\,B_l$ and $E_l^{(s)},\,B_l^{(s)}$ are introduced
by relations analogous to~(\ref{eq:DFS32WUSS}), the second line
of~(\ref{eq:H7TRJS4DZ}) may be written in the form
\begin{equation}
  \label{eq:DLH3KJ7GNG}
\int \mathrm{d}^{3} x \rho
\bigg[
\big(
e\vec{E}+\frac{e}{c}\vec{v} \times \vec{B}
\big)_l
+
\big(
e\vec{E^{(s)}}+\frac{e}{c}\vec{v} \times \vec{B^{(s)}}
\big)_l
\bigg]
\mbox{,}
\end{equation}
which shows that both types of fields, the external fields as well
as the internal fields due to $\vartheta,\,\varphi$, enter the
theory in the same way, namely in the form of a Lorentz force.

The first, externally controlled Lorentz force in~(\ref{eq:DLH3KJ7GNG})
may be eliminated in exactly the same manner as in
section~\ref{sec:gaugecoupling} by writing
\begin{equation}
  \label{eq:DLI3LDF9SNG}
\overline{F^{(T)}_{l}(x,\,p,\,t)}=
\int \mathrm{d}^{3} x \rho
\big(
e\vec{E}+\frac{e}{c}\vec{v} \times \vec{B}
\big)_l
+\int \mathrm{d}^{3} x \rho F^{(I)}_{l}(x,\,p,\,t)
\mbox{.}
\end{equation}
This means that one of the forces acting on the system as a whole is
again given by a Lorentz force; there may be other nontrivial
forces~$F^{(I)}$ which are still to be determined. The second 'internal'
Lorentz force in~(\ref{eq:DLH3KJ7GNG}) can, of course, not be eliminated
in this way. In order to proceed, the third statistical
condition~(\ref{eq:ANA35EQUT}) must be implemented.
To do that it is useful to rewrite Eq.~(\ref{eq:H7TRJS4DZ}) in the form
\begin{equation}
  \label{eq:H23DASU8Z}
\begin{split}
&-\int \mathrm{d}^{3} x
\frac{\partial \rho}{\partial x_{l}}
\bigg[
\bigg( \frac{\partial S}{\partial t}+e \hat{\phi} \bigg)
+\frac{1}{2m}
\sum_{j}
\bigg(
\frac{\partial S}{\partial x_{j}} - \frac{e}{c} \hat{A_j}
\bigg)^{2}
\bigg]\\
&+\int \mathrm{d}^{3} x
\frac{\hbar}{2} \rho  \sin \vartheta
\bigg(
\frac{\partial \vartheta}{\partial x_{l}}
\bigg[
 \frac{\partial \varphi}{\partial t}
+ v_j\frac{\partial \varphi}{\partial x_{j}}
\bigg]
-\frac{\partial \varphi}{\partial x_{l}}
\bigg[
 \frac{\partial \vartheta}{\partial t}
+ v_j\frac{\partial \vartheta}{\partial x_{j}}
\bigg]
\bigg) \\
&= \overline{F^{(I)}_{l}(x,\,p,\,t)}
=\int \mathrm{d}^{3} x \rho F^{(I)}_{l}(x,\,p,\,t)
\mbox{,}
\end{split}
\end{equation}
using~(\ref{eq:DLH3KJ7GNG}),~(\ref{eq:DLI3LDF9SNG}) and the
definition~(\ref{eq:HIDV32PESSF}) of the fields $A_{l}^{(s)}$
and $\phi^{(s)}$.

We interpret the fields $\varphi$ and $\vartheta$ as angles
(with $\varphi$ measured from the $y-$axis of our coordinate system)
determining the direction of a vector
\begin{equation}
  \label{eq:UF27VWN8B}
\vec{s}=\frac{\hbar}{2}
\big(
  \sin \vartheta \sin \varphi \, \vec{e}_x
+ \sin \vartheta \cos \varphi \, \vec{e}_y
+ \cos \vartheta \, \vec{e}_z
\big)
\mbox{,}
\end{equation}
of constant length $\frac{\hbar}{2}$. As a consequence, $\dot{\vec{s}}$
and $\vec{s}$ are perpendicular to each other and the classical force
$\vec{F}^{(R)}$ in Eq.~(\ref{eq:ANA35EQUT}) should be of the form
$\vec{D}\times \vec{s}$, where $\vec{D}$ is an unknown field. In
contrast to the 'external force', we are unable to determine the complete
form of this 'internal' force from the statistical constraint [an alternative
treatment will be reported in section~\ref{sec:spin-as-gauge}] and set
\begin{equation}
  \label{eq:NG89NWG2EN}
\vec{F}^{(R)}=
-\frac{e}{mc}\vec{B} \times \vec{s}
\mbox{,}
\end{equation}
where $\vec{B}$ is the external 'magnetic field', as defined by
Eq.~(\ref{eq:DFS32WUSS}), and the factor in front of $\vec{B}$
has been chosen to yield the correct $g-$factor of the electron.

The differential equation
\begin{equation}
  \label{eq:BBHA87EWUT}
\frac{\mathrm{d}}{\mathrm{d}t} \vec{s}  =
-\frac{e}{mc}\vec{B} \times \vec{s}
\mbox{}
\end{equation}
for particle variables $ \vartheta(t),\;\varphi(t)$
describes the rotational state of a classical magnetic dipole in
a magnetic field, see~\cite{schiller:spinning}. Recall that we do
\emph{not} require that~\eqref{eq:BBHA87EWUT} is fulfilled
in the present theory. The present variables are the fields
$ \vartheta(x,t),\;\varphi(x,t)$ which may be thought of as
describing a kind of 'rotational state' of the statistical ensemble
as a whole, and have to fulfill the 'averaged
version'~(\ref{eq:ANA35EQUT}) of~(\ref{eq:BBHA87EWUT}).

Performing steps similar to the ones described in I (see also
section~\ref{sec:calcwpt}), the third statistical
condition~(\ref{eq:ANA35EQUT}) implies the following
differential relations,
\begin{eqnarray}
\dot{\varphi}+v_j\frac{\partial \varphi}{\partial x_j} & = &
\frac{e}{mc} \frac{1}{\sin \vartheta}
\left(
B_z \sin \vartheta -  B_y \cos \vartheta \cos\varphi
-  B_x \cos \vartheta \sin\varphi
\right) \nonumber  \\
& + & \frac{\cos \varphi }{\sin \vartheta} G_1-
\frac{\sin \varphi }{\sin \vartheta} G_2, \label{eq:Q29D3SJUE} \\
\dot{\vartheta}+v_j\frac{\partial \vartheta}{\partial x_j} & = &
\frac{e}{mc} \left( B_x \cos \varphi  -  B_y \sin\varphi \right)
-\frac{G_3}{\sin \vartheta}
\label{eq:Q29D3SJ22}
\mbox{,}
\end{eqnarray}
for the dynamic variables $\vartheta$ and $\varphi$. These
equations contain three fields $G_i(x,t),\;i=1,2,3$ which
have to obey the conditions
\begin{equation}
  \label{eq:DR4ZGC98FG}
\int \mathrm{d}^{3} x \rho \, G_i =0,\hspace{0.5cm}
\vec{G}\vec{s}=0
\mbox{,}
\end{equation}
and are otherwise arbitrary. The 'total derivatives' of $\varphi$ and
$\vartheta$ in~(\ref{eq:H23DASU8Z}) may now be eliminated with the
help of~(\ref{eq:Q29D3SJUE}),(\ref{eq:Q29D3SJ22}) and the second
line of Eq.~(\ref{eq:H23DASU8Z}) takes the form
\begin{equation}
  \label{eq:AT2UZ5WE8Q}
 \begin{split}
 &\int \mathrm{d}^{3} x \frac{\partial \rho}{\partial x_{l}}
 \frac{e}{mc}s_jB_j+
 \int \mathrm{d}^{3} x \rho \,\frac{e}{mc}
 s_j\frac{\partial}{\partial x_l} B_j \\
 +&\int \mathrm{d}^{3} x \rho \,\frac{\hbar}{2}
 \big(
   \cos\varphi \frac{\partial \vartheta}{\partial x_l} G_1
 - \sin\varphi \frac{\partial \vartheta}{\partial x_l} G_2
 + \frac{\partial \varphi}{\partial x_l} G_3
 \big)
 \mbox{.}
\end{split}
\end{equation}

The second term in~(\ref{eq:AT2UZ5WE8Q}) presents an external
macroscopic force. It may be eliminated from~(\ref{eq:H23DASU8Z}) by
writing
\begin{equation}
  \label{eq:DSK1NW4DNG}
\overline{F^{(I)}_{l}(x,\,p,\,t)}=
\int \mathrm{d}^{3} x \rho
\big(-\mu_j \frac{\partial}{\partial x_l} B_j)
+\overline{F^{(V)}_{l}(x,\,p,\,t)}
\mbox{,}
\end{equation}
where the magnetic moment of the electron $\mu_i=-(e/mc)s_i$ has been
introduced. The first term on the r.h.s. of~(\ref{eq:DSK1NW4DNG})
is the expectation value of the well-known electrodynamical force
exerted by an inhomogeneous magnetic field on the translational
motion of a magnetic dipole; this classical force plays an important
role in the standard interpretation of the quantum-mechanical
Stern-Gerlach effect. It is satisfying that both translational
forces, the Lorentz force as well as this dipole force, can be
derived in the present approach. The remaining unknown force
$\vec{F}^{(V)}$ in~(\ref{eq:DSK1NW4DNG}) leads (in the same way
as in section~\ref{sec:gaugecoupling}) to a mechanical potential
$V$, which will be omitted for brevity.

The integrand of the first term in~(\ref{eq:AT2UZ5WE8Q}) is linear in
the derivative of $\rho$ with respect to $x_l$. It may consequently be
added to the first line of~(\ref{eq:H23DASU8Z}) which has the same
structure. Therefore, it represents (see below) a contribution to
the generalized Hamilton-Jacobi differential equation. The third
term in~(\ref{eq:AT2UZ5WE8Q}) has the mathematical structure of
a force term, but does not contain any externally controlled
fields. Thus, it must also represent a contribution to the generalized
Hamilton-Jacobi equation. This implies that this third term can be
written as
\begin{equation}
  \label{eq:SUM44HJE9W}
\int \mathrm{d}^{3} x \rho \,\frac{\hbar}{2}
 \big(
   \cos\varphi \frac{\partial \vartheta}{\partial x_l} G_1
 - \sin\varphi \frac{\partial \vartheta}{\partial x_l} G_2
 + \frac{\partial \varphi}{\partial x_l} G_3
 \big)=
\int \mathrm{d}^{3} x \frac{\partial \rho}{\partial x_{l}}
 L_0^{\prime}
\mbox{,}
\end{equation}
where $L_0^{\prime}$ is an unknown field depending on $G_1,\,G_2,\,G_3$.

Collecting terms and restricting ourselves, as in
section~\ref{sec:fisher-information}, to an isotropic
law, the statistical condition~\eqref{eq:H23DASU8Z} takes
the form of a generalized Hamilton-Jacobi equation:
\begin{equation}
  \label{eq:DZ3GANU7EHJ}
\bar{L} := \bigg( \frac{\partial S}{\partial t}+e \hat{\phi} \bigg)
+\frac{1}{2m}
\sum_{j}
\bigg(
\frac{\partial S}{\partial x_{j}} - \frac{e}{c} \hat{A_j}
\bigg)^{2}
+\mu_i B_i=L_0
\mbox{.}
\end{equation}
The unknown function $L_0$ must contain $L_0^{\prime}$ but may also
contain other terms, let us write $L_0=L_0^{\prime}+\Delta L_0$.

\section{'Missing' quantum spin terms from Fisher information}
\label{sec:spin-fish-inform}
Let us summarize at this point what has been achieved so far.
We have four coupled differential equations for our dynamic
field variables $\rho,\,S,\,\vartheta,\,\varphi$. The first
of these is the continuity equation~\eqref{eq:contspin}, which
is given, in terms of the present variables, by
\begin{equation}
  \label{eq:CO3SPI4NVAR}
\frac{\partial \rho}{\partial t}+
\frac{\partial}{\partial x_l}
\bigg[
\frac{\rho}{m}
\big(
\frac{\partial S}{\partial x_{j}} - \frac{e}{c} \hat{A_j}
\big)
\bigg]=0
\mbox{.}
\end{equation}
The three other differential equations, the evolution
equations~\eqref{eq:Q29D3SJUE},~\eqref{eq:Q29D3SJ22}
and the generalized Hamilton-Jacobi equation~\eqref{eq:DZ3GANU7EHJ},
do not yet possess a definite mathematical form. They contain four
unknown functions $G_i,\,L_0$ which are constrained, but not
determined, by~\eqref{eq:DR4ZGC98FG},~\eqref{eq:SUM44HJE9W}.

The simplest choice, from a formal point of view, is $G_i=L_0=0$.
In this limit the present theory agrees with Schiller's
field-theoretic (Hamilton-Jacobi) version, see ~\cite{schiller:spinning},
of the equations of motion of a classical dipole. This is a  classical
(statistical) theory despite the fact that it contains
[see~\eqref{eq:HIDV32PESSF}] a number $\hbar$. But this classical
theory is not realized in nature; at least not in the microscopic
domain. The reason is that the simplest choice from a formal point
of view is not the simplest choice from a physical point of view.
The postulate of maximal simplicity (Ockham's razor) implies
equal probabilities and the principle of maximal entropy in classical
statistical physics. A similar principle which is able to 'explain'
the nonexistence of classical physics (in the microscopic domain) is
the principle of minimal Fisher information~\cite{frieden:sciencefisher}.
The relation between the two (classical and quantum-mechanical) principles
has been discussed in detail in I.

The mathematical formulation of the principle of minimal Fisher
information for the present problem requires a generalization,
as compared to I, because we have now several fields with coupled
time-evolution equations. As a consequence, the spatial integral
(spatial average) over $\rho (\bar{L}-L_{0})$  in the variational
problem~(\ref{eq:EEFFRRWXYZ2}) should be replaced by a space-time
integral, and the variation should be performed with respect to
all four variables. The problem can be written in the form
\begin{gather}
\delta \int \mathrm{d} t \int \mathrm{d}^{3} x   \rho
\left( \bar{L} - L_{0} \right) = 0 \label{eq:HH43TRT9RQ} \\
E_a=0,\;\;\;a=S,\,\rho,\,\vartheta,\,\varphi \label{eq:HQ3TR4JWQ}
\mbox{,}
\end{gather}
where $E_a=0$ is a shorthand notation for the
equations~\eqref{eq:CO3SPI4NVAR},~\eqref{eq:DZ3GANU7EHJ},~\eqref{eq:Q29D3SJ22}~\eqref{eq:Q29D3SJUE}. Eqs.~(\ref{eq:HH43TRT9RQ}),~(\ref{eq:HQ3TR4JWQ})
require that the four Euler-Lagrange equations of the variational
problem~(\ref{eq:HH43TRT9RQ}) agree with the differential
equations~(\ref{eq:HQ3TR4JWQ}). This imposes conditions for the
unknown functions $L_{0},\,G_i$. If the \emph{solutions}
of~(\ref{eq:HH43TRT9RQ}),~(\ref{eq:HQ3TR4JWQ}) for $L_{0},\,G_i$ are
inserted in the variational problem~(\ref{eq:HH43TRT9RQ}), the
four relations~(\ref{eq:HQ3TR4JWQ}) become redundant and
$\rho ( \bar{L} - L_{0} )$ becomes the Lagrangian density of our
problem. Thus, Eqs.~(\ref{eq:HH43TRT9RQ}) and~(\ref{eq:HQ3TR4JWQ})
represent a method to construct a Lagrangian.

We assume a functional form
$L_{0}(\chi_{\alpha},\,\partial_{k}\chi_{\alpha},\,
\partial_{k}\partial_{l}\chi_{\alpha})$,
where $\chi_{\alpha}=\rho,\,\vartheta,\,\varphi$. This means $L_{0}$
does not possess an explicit $x,t$-dependence and does not depend
on $S$ (this would lead to a modification of the continuity equation).
We further assume that $L_{0}$ does not depend on time-derivatives of
$\chi_{\alpha}$ (the basic structure of the time-evolution equations
should not be affected) and on spatial derivatives higher than second
order. These second order derivatives must be taken into account but
should not give contributions to the variational equations (a more
detailed discussion of the last point has been given in I).

The variation with respect to $S$ reproduces the continuity
equation which is unimportant for the determination of $L_{0},\,G_i$.
Performing the variation with respect to $\rho,\,\vartheta,\,\varphi$
and taking the corresponding
conditions~\eqref{eq:DZ3GANU7EHJ},~\eqref{eq:Q29D3SJ22}~\eqref{eq:Q29D3SJUE}
into account leads to the following differential equations for
$L_{0},\,G_1\cos \varphi - G_2\sin \varphi $ and $G_3$,
\begin{gather}
-\frac{\partial}{\partial x_{k}}  \frac{\partial}{\partial x_{i}}
\frac{\partial \rho L_0}
{\partial
\frac{\partial^{2} \rho}{\partial x_{k} \partial x_{i}}}
+\frac{\partial}{\partial x_{k}}
\frac{\partial \rho L_0}
{\partial
\frac{\partial \rho}{\partial x_{k} }}
-\rho \frac{\partial L_0}{\partial\rho}=0
 \label{eq:DI2DS8RHO} \\
-\frac{\partial}{\partial x_{k}}  \frac{\partial}{\partial x_{i}}
\frac{\partial \rho L_0}
{\partial\
\frac{\partial^{2} \vartheta}{\partial x_{k} \partial x_{i}}}
+\frac{\partial}{\partial x_{k}}
\frac{\partial \rho L_0}
{\partial
\frac{\partial \vartheta}{\partial x_{k} }}
-\frac{\partial \rho L_0}{\partial\vartheta}
-\frac{\hbar \rho}{2}
\left( G_1\cos \varphi - G_2\sin \varphi  \right)
=0
 \label{eq:D6DSV3ARTH}\\
-\frac{\partial}{\partial x_{k}}  \frac{\partial}{\partial x_{i}}
\frac{\partial \rho L_0}
{\partial\
\frac{\partial^{2} \varphi}{\partial x_{k} \partial x_{i}}}
+\frac{\partial}{\partial x_{k}}
\frac{\partial \rho L_0}
{\partial
\frac{\partial \varphi}{\partial x_{k} }}
-\frac{\partial \rho L_0}{\partial\varphi}
-\frac{\hbar }{2} \rho G_3 =0
\label{eq:D9ZSG4PHI}
\mbox{.}
\end{gather}
The variable $S$ does not occur
in~(\ref{eq:DI2DS8RHO})-(\ref{eq:D9ZSG4PHI}) in  agreement with our
assumptions about the form of $L_0$. It is easy to see that a proper
solution (with vanishing variational contributions from the second
order derivatives) of~(\ref{eq:DI2DS8RHO})-(\ref{eq:D9ZSG4PHI}) is
given by
\begin{gather}
L_0 = \frac{\hbar^{2}}{2m} \bigg[
\frac{1}{\sqrt{\rho}}
\frac{\partial}{\partial \vec{x}} \frac{\partial}{\partial \vec{x}}
\sqrt{\rho}
-\frac{1}{4}\sin^{2} \vartheta
\left( \frac{\partial \varphi}{\partial \vec{x}} \right)^{2}
-\frac{1}{4}
\left( \frac{\partial \vartheta}{\partial \vec{x}} \right)^{2}
\bigg]
 \label{eq:DI3SOLL0HO} \\
\hbar G_1\cos \varphi - \hbar G_2\sin \varphi=
\frac{\hbar^{2}}{2m} \bigg[
\frac{1}{2}\sin 2\vartheta
\left( \frac{\partial \varphi}{\partial \vec{x}} \right)^{2}
-\frac{1}{\rho}
\frac{\partial}{\partial \vec{x}} \rho
\frac{\partial \vartheta}{\partial \vec{x}}
\bigg]
 \label{eq:D8SOLG1G2H}\\
\hbar G_3 = - \frac{\hbar^{2}}{2m} \frac{1}{\rho}
\frac{\partial}{\partial \vec{x}}
\big(
\rho \sin^{2} \vartheta \frac{\partial \varphi}{\partial \vec{x}}
\big)
\label{eq:D9SOLG3U2W}
\mbox{.}
\end{gather}
A new adjustable parameter appears on the r.h.s
of~(\ref{eq:DI3SOLL0HO})-~(\ref{eq:D9SOLG3U2W}) which has
been identified with $\hbar^{2}/2m$, where $\hbar$ is again Planck's
constant. This second  $\hbar$ is related to the quantum-mechanical
principle of maximal disorder. It is in the present approach not
related in any obvious way to the previous "classical" $\hbar$
which denotes the amplitude of a rotation; compare, however, the
alternative derivation of spin in section~\ref{sec:spin-as-gauge}.

The solutions for $G_1,\,G_2$ may be obtained with the
help of the second condition ($\vec{G} \vec{s} =0 $) listed in
Eq.~(\ref{eq:DR4ZGC98FG}). The result may be written in the form
 \begin{equation}
  \label{eq:HRG1G2U4Z}
\begin{split}
G_1 &= \frac{\hbar}{2m} \frac{1}{\rho}
\frac{\partial}{\partial \vec{x}} \rho
\bigg( \frac{1}{2} \sin 2\vartheta \sin \varphi
\frac{\partial \varphi}{\partial \vec{x}}
- \cos \varphi \frac{\partial
\vartheta}{\partial \vec{x}}
\bigg)\\
G_2 &= \frac{\hbar}{2m} \frac{1}{\rho}
\frac{\partial}{\partial \vec{x}} \rho
\bigg( \frac{1}{2} \sin 2\vartheta \cos \varphi
\frac{\partial \varphi}{\partial \vec{x}}
+ \sin \varphi \frac{\partial
\vartheta}{\partial \vec{x}}
\bigg)
\mbox{.}
\end{split}
\end{equation}
Eqs.~(\ref{eq:D9SOLG3U2W}) and~(\ref{eq:HRG1G2U4Z}) show that the
first condition listed in~(\ref{eq:DR4ZGC98FG}) is also satisfied.
The last condition is also fulfilled: $L_0$ can be written as
$L_0^{\prime}+\Delta L_0$, where
\begin{equation}
  \label{eq:AE98VZVDEF}
L_0^{\prime} =
-\frac{\hbar^{2}}{8m} \bigg[
\sin^{2} \vartheta
\left( \frac{\partial \varphi}{\partial \vec{x}} \right)^{2}
+\left( \frac{\partial \vartheta}{\partial \vec{x}} \right)^{2}
\bigg],\;\;\;
\Delta L_0 =
\frac{\hbar^{2}}{2m}
\frac{1}{\sqrt{\rho}}
\frac{\partial}{\partial \vec{x}} \frac{\partial}{\partial \vec{x}}
\sqrt{\rho}
\mbox{,}
\end{equation}
and $L_0^{\prime}$ fulfills~(\ref{eq:SUM44HJE9W}).
We see that $L_0^{\prime}$ is a quantum-mechanical contribution to
the rotational motion while $\Delta L_0$ is related to the probability
density of the ensemble (as could have been guessed considering the
mathematical form of these terms). The last term is the same as in
the spinless case [see~(\ref{eq:QHJH34MF})].

The remaining task is to show that the above solution for $L_0$
does indeed lead to a (appropriately generalized) Fisher functional.
This can be done in several ways. The simplest is to use the following
result due to~\cite{reginatto:pauli}:
\begin{gather}
\int \mathrm{d}^{3} x  \left( -\rho L_{0} \right) =
\frac{\hbar^{2}}{8m} \sum_{j=1}^{3}
\int \mathrm{d}^{3} x\,\sum_{k=1}^{3}
\frac{1}{\rho^{(j)}}
\left(\frac{\partial \rho^{(j)}}{\partial x_k} \right)^{2} \mbox{,}
\label{eq:RW2QI7U9Z}  \\
\rho^{(1)} :=  \rho \sin^{2}\frac{\vartheta}{2}
\cos^{2}\frac{\varphi}{2},\;\;
\rho^{(2)} :=  \rho \sin^{2}\frac{\vartheta}{2}
\sin^{2}\frac{\varphi}{2},\;\;
\rho^{(3)} :=  \rho \cos^{2}\frac{\vartheta}{2}
\label{eq:HJLLH3LKL}
\mbox{.}
\end{gather}
The functions $\rho^{(j)}$  represent the probability that a particle is at
space-time point $x,\,t$ and $\vec{s}$ points into direction $j$.
Inserting~(\ref{eq:DI3SOLL0HO}) the validity of~(\ref{eq:RW2QI7U9Z})
may easily be verified. The r.h.s. of Eq.~(\ref{eq:RW2QI7U9Z}) shows
that the averaged value of $L_{0}$ represents indeed a Fisher functional,
which completes our calculation of the 'quantum terms' $L_{0},\,G_i$.

Summarizing, our assumption, that under certain external conditions
four state variables instead of two may be required, led to a
nontrivial result, namely the four coupled
differential equations~\eqref{eq:CO3SPI4NVAR},~\eqref{eq:DZ3GANU7EHJ},~\eqref{eq:Q29D3SJ22},~\eqref{eq:Q29D3SJUE}
with $L_{0},\,G_i$ given
by~(\ref{eq:DI3SOLL0HO}),~(\ref{eq:HRG1G2U4Z}),~(\ref{eq:D9SOLG3U2W}).
The external condition which stimulates this splitting is given
by a gauge field; the most important case is a magnetic
field $\vec{B}$ but other possibilities do exist (see below).
These four differential equations are equivalent to the much simpler
differential equation
\begin{equation}
  \label{eq:DT26RPAULI}
 \big(\frac{\hbar}{\imath} \frac{\partial}{\partial t}
+e\phi \big) \hat{\psi}
+ \frac{1}{2 m }
\big(\frac{\hbar}{\imath} \frac{\partial}{\partial \vec{x}} -
 \frac{e}{c} \vec{A} \big)^{2} \hat{\psi}
+ \mu_{B}\vec{\sigma}\vec{B} \hat{\psi}=0
\mbox{,}
\end{equation}
which is linear in the complex-valued two-component state variable
$\hat{\psi}$ and is referred to as Pauli equation (the components of
the vector $\vec{\sigma}$ are the three Pauli matrices and
$\mu_{B}=-e\hbar/2mc$). To see the equivalence one
writes, see~\cite{takabayasi:vector},~\cite{holland:quantum},
\begin{equation}
  \label{eq:ASF7UE1PSIHT}
\hat{\psi}=\sqrt{\rho}\, \mathrm{e}^{\frac{\imath}{\hbar}S}
\left(
\begin{array}{cc}
\cos \frac{\vartheta}{2} \mathrm{e}^{\imath \frac{\varphi}{2}}  & \\
\\
\imath \sin \frac{\vartheta}{2} \mathrm{e}^{-\imath \frac{\varphi}{2}} &
\end{array}
\right)
\mbox{,}
\end{equation}
and evaluates the real and imaginary parts of the two scalar
equations~(\ref{eq:DT26RPAULI}). This leads to the four differential
equations~\eqref{eq:CO3SPI4NVAR},~\eqref{eq:DZ3GANU7EHJ},~\eqref{eq:Q29D3SJ22}~\eqref{eq:Q29D3SJUE} and completes the present spin theory.

In terms of the real-valued functions $\rho,\,S,\,\vartheta,\,\varphi$
the quantum-mechanical
solutions~(\ref{eq:DI3SOLL0HO}),~(\ref{eq:D9SOLG3U2W}),~(\ref{eq:HRG1G2U4Z})
for $L_{0},\,G_i$ look complicated in comparison to the classical solutions
$L_{0}=0,\,G_i=0$. In terms of the variable $\hat{\psi}$ the situation changes
to the contrary: The quantum-mechanical equation becomes simple (linear)
and the classical equation, which has been derived
by~\cite{schiller:spinning}, becomes complicated (nonlinear).
The simplicity of the underlying physical principle (principle of maximal disorder)
leads to a simple mathematical representation of the final basic equation (if
a complex-valued state function is introduced). One may also say that the
linearity of the equations is a consequence of this principle of maximal disorder.
This is the deeper reason why it has been possible, see~\cite{klein:schroedingers}, to derive Schr\"odinger's equation  from a set of assumptions including linearity.

Besides the Pauli equation we found, as a second important result
of our spin calculation, that the following local force is compatible
with the statistical constraint:
\begin{equation}
  \label{eq:LO3DAZI8FF}
\vec{F}^{L}+\vec{F}^{I}=
e \left( \vec{E} + \frac{1}{c} \vec{v} \times \vec{B} \right)
-\vec{\mu} \cdot \frac{\partial}{\partial\vec{x} } \vec{B}
\mbox{.}
\end{equation}
Here, the velocity field $\tilde{\vec{v}}(x,t)$ and the magnetic
moment field $\vec{\mu}(x,t)=-(e/mc)\vec{s}(x,t)$ have been
replaced by corresponding particle quantities $\vec{v}(t)$
and $\vec{\mu}(t)$; the dot denotes the inner product between
$\vec{\mu}$ and $\vec{B}$. The first force in~(\ref{eq:LO3DAZI8FF}),
the Lorentz force, has been derived here from first principles without
any additional assumptions.  The same cannot be said about the second
force which takes this particular form as a consequence of some additional assumptions
concerning the form of the 'internal force' $\vec{F}^{R}$
[see~(\ref{eq:NG89NWG2EN})]. In particular, the field appearing
in $\vec{F}^{R}$ was arbitrary as well as the proportionality
constant (g-factor of the electron) and had to be adjusted by hand.
It is well-known that in a relativistic treatment the spin term appears
automatically if the potentials are introduced. Interestingly, this unity is not
restricted to the relativistic regime. Following~\cite{arunsalam:hamiltonians}
and~\cite{gould:intrinsic} we report in the next section an alternative  (non-relativistic)
derivation of spin, which does not contain any arbitrary fields or constants - but is unable
to yield the expression~\eqref{eq:LO3DAZI8FF} for the macroscopic electromagnetic
forces.

In the present treatment spin has been introduced as a property of
an ensemble and not of individual particles. Similar views may be found in the
literature, see~\cite{ohanian:spin}. Of course, it is difficult to
imagine the properties of an ensemble as being completely independent
from the properties of the particles it is made from. The question whether
or not a property 'spin' can be ascribed to single particles is a subtle
one. Formally, we could assign a probability of being in a state $i$
($i=1,2$) to a particle  just as we assign a probability for being at
a position $\vec{x} \in R^{3}$. But contrary to position, no classical meaning - and
no classical measuring device - can be associated with the discrete degree of
freedom $i$. Experimentally, the measurement of the 'spin of a single electron'
is  - in contrast to the measurement of its position - a notoriously difficult task.
Such experiments, and a number of other interesting questions related to
spin, have been discussed by \cite{morrison.spin}.

\section{Spin as a consequence of a multi-valued phase}
\label{sec:spin-as-gauge}

As shown by~\cite{arunsalam:hamiltonians},~\cite{gould:intrinsic}, and others,
spin in non-relativistic QT may be introduced in exactly the same manner as the
electrodynamic potentials. In this section we shall apply a slightly modified version
of their method and try to derive spin in an alternative way - which avoids the shortcoming
mentioned in the last section.

\cite{arunsalam:hamiltonians} and \cite{gould:intrinsic} introduce the potentials by
applying the well-known minimal-coupling rule to the free Hamiltonian. In the present
treatment this is achieved by making the quantity $S$ multi-valued. The latter approach
seems intuitively preferable considering the physical meaning of the corresponding
classical quantity. Let us first review the essential steps [see~\cite{klein:schroedingers}
for more details]  in the process of creating potentials in the \emph{scalar} Schr{\"o}dinger
equation:
\begin{itemize}
\item Chose  a free Schr{\"o}dinger equation with single-valued state function.
\item 'Turn on' the interaction by making the state function multi-valued
(multiply it with a multi-valued phase factor)
\item Shift the multi-valued phase factor to the left of all differential
operators, creating new terms (potentials) in the differential equation.
\item Skip the multi-valued phase. The final state function is again
single-valued.
\end{itemize}

Let us adapt this method for the derivation of spin (considering spin
one-half only). The first and most important step is the identification of the  free Pauli
equation. An obvious choice is
\begin{equation}
  \label{eq:FREEPAONE}
 \left[ \frac{\hbar}{\imath} \frac{\partial}{\partial t}
+ \frac{1}{2 m }
\big(\frac{\hbar}{\imath} \frac{\partial}{\partial \vec{x}}\big)^{2}
+V \right]  \bar{\psi} =0
\mbox{,}
\end{equation}
where $\bar{\psi}$ is a single-valued \emph{two-component} state
function; (\ref{eq:FREEPAONE}) is essentially a duplicate of Schr{\"o}dinger's
equation. We may of course add arbitrary vanishing terms to the expression
in brackets. This seems trivial, but some of these terms may vanish \emph{only}
if applied to a single-valued $\bar{\psi} $ and may lead to non-vanishing
contributions if applied later (in the second of the above steps) to a multi-valued
state function  $\bar{\psi}^{multi} $.

In order to investigate this possibility, let us rewrite
Eq.~\eqref{eq:FREEPAONE} in the form
\begin{equation}
  \label{eq:FREPZONE2}
 \left[  \hat{p}_{0} + \frac{1}{2 m }\vec{\hat{p}}\vec{\hat{p}}
+V \right]  \bar{\psi} =0
\mbox{,}
\end{equation}
where $\hat{p}_{0}$  is an abbreviation for the first term
of~\eqref{eq:FREEPAONE} and the spatial derivatives are given by
\begin{equation}
  \label{eq:FJU67LE2}
\vec{\hat{p}}= \hat{p}_{k}\vec{e}_{k},\hspace{1cm}\hat{p}_{k}=
\frac{\hbar}{\imath} \frac{\partial}{\partial x_{k}}
\mbox{.}
\end{equation}
All terms in the bracket in~(\ref{eq:FREPZONE2}) are to be multiplied
with a $2x2$ unit-matrix $E$ which has not be written down. Replace
now the derivatives in~(\ref{eq:FREPZONE2}) according to
\begin{equation}
  \label{eq:REPLW344N}
\hat{p}_{0} \Rightarrow \hat{p}_{0}M_{0},\hspace{1cm}
\vec{\hat{p}} \Rightarrow \vec{\hat{p}}_{k} M_{k}
\mbox{,}
\end{equation}
where $M_{0},\,M_{k}$ are hermitian $2x2$ matrices with constant coefficients,
which should be constructed in such a way that the new equation agrees
with~(\ref{eq:FREPZONE2}) for single-valued $\bar{\psi}$, i.e. assuming
the validity of the condition
\begin{equation}
  \label{eq:ESBNN3HG}
\left( \hat{p}_{i}\hat{p}_{k} - \hat{p}_{k}\hat{p}_{i}\right) \bar{\psi} =0
\mbox{.}
\end{equation}
This leads to the condition
\begin{equation}
  \label{eq:AN2CFO8M}
M_{0}^{-1}M_{i}M_{k}=E \delta_{ik}+T_{ik}
\mbox{,}
\end{equation}
where $T_{ik}$ is a $2x2$ matrix with two cartesian indices $i,\,k$, which
obeys $T_{ik}=-T_{ki}$. A solution of~(\ref{eq:AN2CFO8M}) is given by
$M_{0}=\sigma_{0},\,M_{i}=\sigma_{i}$, where $\sigma_{0},\,\sigma_{i}$ are
the four Pauli matrices. In terms of this solution, Eq.~(\ref{eq:AN2CFO8M}) takes the form
\begin{equation}
  \label{eq:HW2CF78M}
\sigma_{i}\sigma_{k}=\sigma_{0}\delta_{ik}+\imath \varepsilon_{ikl}\sigma_{l}
\mbox{.}
\end{equation}
Thus, an alternative free Pauli-equation, besides~(\ref{eq:FREEPAONE}) is
given by
\begin{equation}
  \label{eq:FEE7PAUDZW}
 \left[ \frac{\hbar}{\imath} \frac{\partial}{\partial t}
+ \frac{1}{2 m }
\left(\frac{\hbar}{\imath}\right)^{2}
\sigma_{i} \frac{\partial}{\partial x_{i}}
\sigma_{k} \frac{\partial}{\partial x_{k}}
+V \right]  \bar{\psi} =0
\mbox{.}
\end{equation}
The quantity in the bracket is the generalized Hamiltonian constructed
by~\cite{arunsalam:hamiltonians} and~\cite{gould:intrinsic}. In the present approach
gauge fields are introduced by means of a multi-valued phase. This leads to the same
formal consequences as the minimal coupling rule but allows us to conclude that the
second free Pauli equation~(\ref{eq:FEE7PAUDZW}) is \emph{more appropriate} than
the first, Eq.~(\ref{eq:FREEPAONE}), because it is more general with regard to the
consequences of multi-valuedness. This greater generality is due to the presence of
the second term on the r.h.s. of~(\ref{eq:HW2CF78M}).

The second step is to turn on the multi-valuedness in Eq.~(\ref{eq:FEE7PAUDZW}),
$\bar{\psi} \Rightarrow \bar{\psi}^{multi}$, by multiplying $\bar{\psi}$  with a
multi-valued two-by-two matrix. This matrix must be chosen in such a way that the
remaining steps listed above lead to Pauli's equation~(\ref{eq:DT26RPAULI})
in presence of an gauge field. Since in our case the final result~(\ref{eq:DT26RPAULI})
is known, this matrix may be found by performing the inverse process, i.e. performing
a singular gauge transformation $\hat{\psi}=\Gamma \bar{\psi}^{multi}$ of Pauli's
equation~(\ref{eq:DT26RPAULI})  from $\hat{\psi}$ to $\bar{\psi}^{multi}$, which
\emph{removes} all electrodynamic terms from~(\ref{eq:DT26RPAULI}) and
creates Eq.~(\ref{eq:FEE7PAUDZW}). The final result for the matrix $\Gamma$ is given by
\begin{equation}
  \label{eq:TFRF34TMG}
\Gamma=E \exp
\bigg\{
\imath \frac{e}{\hbar c}
\int^{x,t} \left[\mathrm{d}x_{k}' A_k(x',t')-c\mathrm{d}t' \phi(x',t')\right]
 \bigg\}
\mbox{,}
\end{equation}
and agrees, apart from the unit matrix $E$, with the multi-valued factor
introduced previously [see~\eqref{eq:AWIZ7MMVF} and~\eqref{eq:HDIS3JWUI}]
leading to the electrodynamic potentials. The inverse transition from~(\ref{eq:FEE7PAUDZW})
to~(\ref{eq:DT26RPAULI}), i.e. the creation of the potentials and the Zeeman term, can be
performed by using the inverse of~\eqref{eq:TFRF34TMG}.

The Hamiltonian~(\ref{eq:FEE7PAUDZW}) derived by~\cite{arunsalam:hamiltonians}
and~\cite{gould:intrinsic} shows
that spin can be described by means of the same abelian gauge theory that leads to the
standard quantum mechanical gauge coupling terms; no new adjustable fields or parameters appear.
The only requirement is that the appropriate free Pauli equation~\eqref{eq:FEE7PAUDZW} is chosen as
starting point. The theory of~\cite{dartora_cabrera:magnetization}, on the other hand, started
from the alternative (from the present point of view inappropriate) free Pauli equation~\eqref{eq:FREEPAONE} and
leads to  the conclusion that spin must be described by a non-abelian gauge theory.

As far as our derivation of non-relativistic QT is concerned we have now two alternative,
and in a sense complementary, possibilities to introduce spin. The
essential step in the second (Arunsalam-Gould) method is the transition
from~\eqref{eq:FREEPAONE} to the equivalent free Pauli equation~\eqref{eq:FEE7PAUDZW}.
This step is a remarkable short-cut for the complicated calculations,
performed in the last section, leading to
the various terms required by the principle of minimal Fisher information. The Arunsalam-Gould
method is unable to provide the shape~\eqref{eq:LO3DAZI8FF} of the corresponding
macroscopic forces but is very powerful insofar as no adjustable quantities are required. It will
be used in the next section to perform the transition to an arbitrary number of particles.

\section{Transition to N particles as final step to non-relativistic
quantum theory}
\label{sec:final-step-to}
In this section the present derivation of non-relativistic QT is completed by
deriving Schr\"odinger's  equation for an arbitrary number $N$ of particles or, more precisely, for statistical ensembles of identically prepared experimental
arrangements involving $N$ particles.

In order to generalize the results of sections~\ref{sec:calcwpt} and~\ref{sec:fisher-information},
a convenient set of $n=3N$ coordinates $q_1,...q_n$ and masses $m_1,...m_n$ is defined by
\begin{equation}
\label{eq:T38AN5OD}
\begin{split}
\left( q_{1},q_{2},q_{3},...q_{n-2},q_{n-1},q_{n}\right)&=
\left(x_1,y_1,z_1,...,x_N,y_N,z_N \right) \mbox{,}
\\
\left( m_{1},m_{2},m_{3},...m_{n-2},m_{n-1},m_{n}\right) &=
\left(m_1,m_1,m_1,...,m_N,m_N,m_N \right)\mbox{.}
\end{split}
\end{equation}
The index $I=1,...N$ is used to distinguish particles, while indices $i,k,..$ are used here to distinguish
the $3N$ coordinates $q_1,...q_n$.   No new symbol has been introduced in~\eqref{eq:T38AN5OD}
to distinguish the masses $m_I$ and $m_i$ since there is no danger of confusion in anyone of the
formulas below. However, the indices of masses will be frequently written in the form $m_{(i)}$ in
order to avoid ambiguities with regard to the summation convention. The symbol $Q$ in arguments
denotes dependence on all $q_1,...q_n$. In order to generalize the results of section~\ref{sec:spin-as-gauge}
a notation $x_{I,k}$, $\vec{x}_{I}$, and $m_{I}$ (with $I=1,...,N$ and $k=1,2,3$) for coordinates,
positions, and masses will be more convenient.

The basic relations of section~\ref{sec:calcwpt}, generalized in an obvious way to
$N$ particles, take the form
\begin{gather}
\frac{\partial \rho(Q,t)}{\partial t}+
\frac{\partial}{\partial q_k} \frac{\rho(Q,t)}{m_{(k)}}\frac{\partial
S(Q,t)}{\partial q_k}=0
\label{eq:CEQ34FNPT}\\
\frac{\mathrm{d}}{\mathrm{d}t} \overline{q_k}  =
\frac{1}{m_{(k)}}\overline{p_k}
\label{eq:FIRNAT12HL}\\
\frac{\mathrm{d}}{\mathrm{d}t} \overline{p_k}  =
\overline{F_k(Q,t)}
\label{eq:SE28UA5SHL}\\
\overline{q_k} =  \int \mathrm{d} Q\, \rho(Q,\,t)\, q_k
\label{eq:UZ1K7ZTWT}
\end{gather}
Here,  $S$ is a single-valued variable; the multi-valuedness will be added later, following the method of section~\ref{sec:spin-as-gauge}.

The following calculations may be performed in complete analogy to the
corresponding steps of section~\ref{sec:calcwpt}. For the present $N-$dimensional
problem, the vanishing of the surface integrals, occurring in the course of various partial
integrations, requires that $\rho$ vanishes exponentially in arbitrary directions of the
configuration space.  The final conclusion to be drawn from
Eqs.~\eqref{eq:CEQ34FNPT}-~\eqref{eq:UZ1K7ZTWT} takes the form
\begin{equation}
  \label{eq:AUBGJJV9F}
\sum_{j=1}^{n} \frac{1}{2m_{(j)}}  \left(\frac{\partial S}{\partial q_{j}} \right)^{2}+
\frac{\partial S}{\partial t} + V = L_0,\hspace{1cm}
\int \mathrm{d}Q\,\frac{\partial\rho}{\partial q_{k}}\,L_{0}= 0
\mbox{,}
\end{equation}
The remaining problem is the determination of the unknown function $L_0$, whose
form is constrained by the condition defined in Eq.~\eqref{eq:AUBGJJV9F}.

$L_0$ can be determined using again the principle of minimal Fisher information,
see I for details. Its implementation in the present framework takes the form
\begin{gather}
\delta \int \mathrm{d} t \int \mathrm{d} Q\, \rho
\left( L - L_{0} \right) = 0 \label{eq:HHWE8AAA} \\
E_a=0,\;\;\;a=S,\,\rho \label{eq:H2WQ45JHJ}
\mbox{,}
\end{gather}
where $E_S=0,\,E_{\rho}=0$ are shorthand notations for the two basic
equations~\eqref{eq:AUBGJJV9F} and~\eqref{eq:CEQ34FNPT}. As before,
Eqs.~\eqref{eq:HHWE8AAA},~\eqref{eq:H2WQ45JHJ} represent a method to
construct a Lagrangian. After determination of  $L_0$ the three relations listed
in~\eqref{eq:HHWE8AAA},~\eqref{eq:H2WQ45JHJ} become redundant
and~\eqref{eq:H2WQ45JHJ} become the fundamental equations of the
$N-$particle theory.

The following calculation can be performed in complete analogy to
the case $N=1$ reported in section~\ref{sec:fisher-information}.
All relations remain valid if the upper summation limit $3$
is replaced by $3N$.  This is also true for the final result, which takes the
form
\begin{equation}
  \label{eq:DES3NR8LL}
L_{0}=\frac{\hbar^{2}}{4\rho}\left[
-\frac{1}{2\rho}
\frac{1}{m_{(j)}}
\frac{\partial \rho}{\partial q_{j}}
\frac{\partial \rho}{\partial q_{j}}
+\frac{1}{m_{(j)}}
\frac{\partial^{2} \rho}
{\partial q_{j}\partial q_{j}}
 \right]
\mbox{.}
\end{equation}
If a complex-valued variable $\psi$, defined as in~(\ref{eq:HDIS3JWUI}), is
introduced, the two basic relations $E_a=0$ may be condensed into the single
differential equation,
\begin{equation}
  \label{eq:DGSFF3R4E}
\Bigg[
 \frac{\hbar}{\imath} \frac{\partial}{\partial t}
+\sum_{I=1}^{N}\frac{1}{2 m_{(I)} }
\left( \frac{\hbar}{\imath}\frac{\partial}{\partial x_{I,k}} \right)
\left( \frac{\hbar}{\imath}\frac{\partial}{\partial x_{I,k}} \right)
+V \Bigg] \psi=0
\mbox{,}
\end{equation}
which is referred to as $N-$particle Schr\"odinger equation, rewritten here in the more familiar
form using particle indices. As is well-known, only approximate solutions of this partial differential
equation of order $3N+1$ exist for realistic systems. The inaccessible complexity of
quantum-mechanical solutions for large $N$ is not reflected in the abstract Hilbert space
structure (which is sometimes believed to characterize the whole of QT) but plays probably a
decisive role for a proper description of the mysterious relation between QT and the
macroscopic world.

Let us now generalize the Arunsalam-Gould method, discussed in section~\ref{sec:spin-as-gauge},
to an arbitrary number of particles. We assume, that the considered  $N-$particle statistical ensemble
responds in $2^{N}$ ways to the external electromagnetic field. This means we restrict ourselves
again, like in section~\ref{sec:spin},~\ref{sec:spin-fish-inform} to spin one-half. Then, the state
function may be written as $\psi(x_1,s_1;....x_I,s_I;...x_N,s_N)$ where $s_I=1,\,2$. In the first of the
steps listed at the beginning of section~\ref{sec:spin-as-gauge}, a differential equation, which is
equivalent to Eq.~\eqref{eq:DGSFF3R4E} for single-valued $\psi$ but may give non-vanishing
contributions for multi-valued $\psi$, has to be constructed.  The proper generalization of Eq.~(\ref{eq:FEE7PAUDZW}) to arbitrary $N$ takes the form
 \begin{equation}
  \label{eq:DSHZ9L7W4E}
\Bigg[
 \frac{\hbar}{\imath} \frac{\partial}{\partial t}
+\sum_{I=1}^{N}\frac{1}{2 m_{(I)} }
\left( \frac{\hbar}{\imath} \sigma_{k}^{(I)} \frac{\partial}{\partial x_{I,k}}  \right)
\left( \frac{\hbar}{\imath} \sigma_{l}^{(I)} \frac{\partial}{\partial x_{I,l}}  \right)
+V \Bigg] \psi=0
\mbox{,}
\end{equation}
where the Pauli matrices $\sigma_{k}^{(I)}$ operate by definition only on the two-dimensional
subspace spanned by the variable $s_{I}$. In the second step we perform the replacement
\begin{equation}
  \label{eq:TF4SDP8HMG}
\psi \Rightarrow
\exp
\bigg\{
-\frac{\imath}{\hbar}
\sum_{I=1}^{N}   \frac{e_{I}}{c}
\sum_{k=1}^{3}
\int^{\vec{x}_{I},t} \left[ \mathrm{d}x_{I,k}' A_k(x_{I}',t')-c\mathrm{d}t' \phi(x_{I}',t')\right]
 \bigg\} \psi
\mbox{,}
\end{equation}
using a multi-valued phase factor, which is an obvious generalization of
Eq.~\eqref{eq:TFRF34TMG}. The remaining  steps, in the listing of
section~\ref{sec:spin-as-gauge}, lead in a straightforward way to the
final result
\begin{equation}
\label{eq:DW2FPGF9NP}
\begin{split}
\Bigg[
 \frac{\hbar}{\imath} \frac{\partial}{\partial t}
&+\sum_{I=1}^{N}e_{I}\phi(x_{I},t)
+\sum_{I=1}^{N}  \sum_{k=1}^{3}    \frac{1}{2 m_{(I)} }
\left(
\frac{\hbar}{\imath}\frac{\partial}{\partial x_{I,k}} -\frac{e_{I}}{c}A_{k}(x_{I},t)
\right)^{2}\\
&+\sum_{I=1}^{N} \mu_{B}^{(I)}\sigma_{k}^{(I)}B_{k}(x_{I},t)
+V(x_1,...,x_N,t) \Bigg] \psi=0
\end{split}
\mbox{,}
\end{equation}
where $\mu_{B}^{(I)}=-\hbar e_{I}/2m_{I}c$ and $\vec{B}=\mathrm{rot}\vec{A}$.
The mechanical potential $V(x_1,...,x_N,t)$ describes a general many-body force
but contains, of course, the usual sum of two-body potentials as a special case.  Eq.~\eqref{eq:DW2FPGF9NP}
is the $N-$body version of Pauli's equation and completes - in the sense
discussed at the very beginning of this paper - the present derivation of
non-relativistic QT.

\section{The classical limit of quantum theory is a statistical theory}
\label{sec:classical-limit}

The classical limit of Schr\"odinger's equation plays an important role for two
topics discussed in the next section, namely the interpretation of QT and the particular significance of potentials in QT; to
study these questions it is sufficient to consider a single-particle ensemble
described by a single state function. This 'classical limit theory' is given by the two differential equations
\begin{gather}
\frac{\partial \rho}{\partial t}+\frac{\partial}{\partial x_{k}}
\frac{\rho}{m} \left(
\frac{\partial S}{\partial x_{k}}-\frac{e}{c} A_{k}
\right)=0
  \label{eq:CONACL7TMF}
\mbox{,} \\
\frac{\partial S}{\partial t}+e\phi + \frac{1}{2m}
\sum_{k}
\left(
\frac{\partial S}{\partial x_{k}}-\frac{e}{c} A_{k}
\right)^{2} +V =0
\label{eq:QHJCL74MF}
\mbox{,}
\end{gather}
which are obtained from Eqs.~(\ref{eq:CONH3TMF})
and~(\ref{eq:QHJH34MF}) by performing the limit $\hbar \to 0$.
The quantum mechanical theory~(\ref{eq:CONH3TMF})
and~(\ref{eq:QHJH34MF}) and the classical theory~(\ref{eq:CONACL7TMF})
and~(\ref{eq:QHJCL74MF}) show fundamentally the same mathematical
structure; both are initial value problems for the variables $S$ and
$\rho$ obeying two partial differential equations. The difference
is the absence of the last term on the l.h.s. of~(\ref{eq:QHJH34MF})
in the corresponding classical equation~(\ref{eq:QHJCL74MF}). This
leads to a \emph{decoupling} of $S$ and $\rho$ in~(\ref{eq:QHJCL74MF});
the identity of the classical object described by $S$ is no longer
affected by statistical aspects described by $\rho$.

The field theory~(\ref{eq:CONACL7TMF}),~(\ref{eq:QHJCL74MF})
for the two 'not decoupled' fields $S$ and $\rho$ is obviously very
different from classical mechanics which is formulated in terms of
trajectories. The fact that one of these equations,
namely~(\ref{eq:QHJCL74MF}), agrees with the Hamilton-Jacobi
equation, does not change the situation
since the presence of the continuity equation~(\ref{eq:CONACL7TMF})
cannot be neglected. On top of that, even if it could be neglected,
Eq.~(\ref{eq:QHJCL74MF}) would still be totally different from
classical mechanics: In order to construct particle trajectories
from the partial differential equation~(\ref{eq:QHJCL74MF}) for the
field $S(x,t)$, a number of clearly defined mathematical manipulations,
which are part of the classical theory of canonical
transformations, see~\cite{greiner:mechanics_systems}, must be performed. The
crucial point is that the latter theory is \emph{not} part of QT and cannot be added
'by hand' in the limit $\hbar \to 0$. Thus,~(\ref{eq:CONACL7TMF}),~(\ref{eq:QHJCL74MF})
is, like QT, an \emph{indeterministic} theory predicting not values of single
event observables but only probabilities, which must be verified
by ensemble measurements.

Given that we found a solution $S(x,t)$, $\rho(x,t)$
of~\eqref{eq:CONACL7TMF},~\eqref{eq:QHJCL74MF} for given initial
values, we may ask which experimental predictions can be made
with the help of these quantities. Using the fields
$\vec{\tilde{p}}(x,\,t)$, $\tilde{E}(x,\,t)$
defined by Eqs.~\eqref{eq:DOJUE8W22},~\eqref{eq:DOJUE8W11}, the Hamilton-Jacobi
equation~\eqref{eq:QHJCL74MF} takes the form
\begin{equation}
  \label{eq:DIE23TFADE}
\frac{\vec{\tilde{p}}^{2}(x,\,t)}{2m}+V(x,t)=\tilde{E}(x,\,t)
\mbox{,}
\end{equation}
The l.h.s. of~\eqref{eq:DIE23TFADE} depends on the
field $\vec{\tilde{p}}$ in the same way as a classical particle
Hamiltonian on the (gauge-invariant) kinetic momentum $\vec{p}$.
We conclude that the field $\vec{\tilde{p}}(x,\,t)$ describes a
mapping from space-time points to particle momenta: If a particle
(in an external electromagnetic field) is found at time $t$ at the
point $x$, then its kinetic momentum is given by
$\vec{\tilde{p}}(x,\,t)$. This is not a deterministic prediction
since we can not predict if a single particle will be or will not
be at time $t$ at point $x$; the present theory gives only a probability
$\rho(x,t)$ for such an event. Combining our findings about
$\vec{\tilde{p}}(x,\,t)$ and $x$ we conclude that the experimental prediction which
can be made with the help of $S(x,t)$, $\rho(x,t)$ is given by the
following phase space probability density:
\begin{equation}
  \label{eq:PPEE44W2S}
w(x,p,t)=\rho(x,t)\delta^{(3)}(p-\frac{\partial
\tilde{S}(x,t)}{\partial x})
\mbox{.}
\end{equation}
Eq.~\eqref{eq:PPEE44W2S} confirms our claim that the classical
limit theory is a statistical theory. The one-dimensional version
of~\eqref{eq:PPEE44W2S} has been obtained before by means of a slightly
different method in I. The deterministic element [realized by the
delta-function shaped probability in~(\ref{eq:PPEE44W2S})] contained
in the classical statistical
theory~(\ref{eq:CONACL7TMF}),~(\ref{eq:QHJCL74MF}) is \emph{absent} in
QT, see I.

Eqs.~(\ref{eq:CONACL7TMF}),~(\ref{eq:QHJCL74MF}) constitute the
mathematically well-defined limit $\hbar \to 0$ of Schr\"odinger's
equation. Insofar as there is general agreement with regard to
two points, namely that (i) 'non-classicality' (whatever this may mean
precisely) is expressed by a nonzero $\hbar$, and that (ii)
Schr\"odinger's equation is the most important relation of quantum
theory, one would also expect general agreement with regard to a further
point, namely that Eqs.~(\ref{eq:CONACL7TMF}),~(\ref{eq:QHJCL74MF})
present essentially (for a three-dimensional configuration space)
the \emph{classical limit} of quantum mechanics. But this is, strangely
enough, not the case. With a few
exceptions, see~\cite{vanvleck:correspondence},~\cite{schiller:quasiclassical},~\cite{ballentine:ehrenfest},~\cite{shirai:reinterpretation},~\cite{klein:schroedingers},
most works (too many to be quoted) take it for granted that the
classical limit of quantum theory is classical mechanics. The objective of
papers like~\cite{rowe:classical},~\cite{werner:classical},~\cite{landau_lj:macroscopic},~\cite{allori_zanghi:classical}
 devoted to ``..the classical limit of quantum mechanics..`` is very often
not the problem: ''\emph{what is} the classical limit of quantum
mechanics ?'' but rather: ``\emph{how to bridge the gap} between
quantum mechanics and classical mechanics ?''. Thus, the fact that
classical mechanics is the classical limit of quantum mechanics is
considered as \emph{evident} and any facts not compatible with it - like
Eqs.~(\ref{eq:CONACL7TMF}),~(\ref{eq:QHJCL74MF}) - are denied.

What, then, is the reason for this widespread denial of reality ?
One of the main reasons is the principle of reductionism which
still rules the thinking of most physicists today. The reductionistic
ideal is a hierarchy of physical theories; better theories have an
enlarged domain of validity and contain 'inferior' theories as
special cases. This principle which has been extremely successful in
the past \emph{dictates} that classical mechanics is a special case of
quantum theory. Successful as this idea might have been during a
long period of time it is not necessarily universally true; quantum
mechanics and classical mechanics describe different domains of
reality, both may be true in their own domains of validity. Many
phenomena in nature indicate that the principle of reductionism
(alone) is insufficient to describe reality, see~\cite{laughlin:everything}.
Releasing ourselves from the metaphysical principle of reductionism,
we accept that the classical limit of quantum mechanics for a
three-dimensional configuration space is the statistical theory defined by
Eqs.~(\ref{eq:CONACL7TMF}),~(\ref{eq:QHJCL74MF}). It is clear
that this theory is not realized in nature (with the same physical
meaning of the variables) because $\hbar$ is different from zero.
But this is a different question and does not affect the conclusion.

\section{Extended discussion}
\label{sec:discussion}

In this paper  it has been shown, continuing the work of I, that the basic differential
equation of non-relativistic QT may be derived from a number of clearly defined assumptions
of a statistical nature. Although this does not exclude the possibility of other derivations, we
consider this success as a strong argument in favor of the statistical interpretation of QT.

This result explains also, at least partly, the success of the canonical quantization
rules~\eqref{eq:DTR43QUPR}. Strictly speaking, these rules have only be derived for a
particular (though very important) special case, the Hamiltonian. However, one
can expect that~\eqref{eq:DTR43QUPR} can be verified for all meaningful physical
observables\footnote{As indicated by preliminary calculations of the angular momentum
relation corresponding to Eq.~\eqref{eq:SECONAETSCHL}}. On the other hand, it cannot
be expected that the rules~\eqref{eq:DTR43QUPR} hold for arbitrary functions of $x,\,p$;
each case has to be investigated separately. Thus, the breakdown of~\eqref{eq:DTR43QUPR},
as expressed by Groenewold's theorem, is no surprise.

The fundamental Ehrenfest-like relations of the present theory establish [like the formal
rules~\eqref{eq:DTR43QUPR}] a \emph{correspondence} between particle mechanics
and QT. Today, philosophical questions concerning, in particular, the 'reality' of particles
play an important role in the thinking of some physicists. So: 'What is this theory about.. ?' While the present author is no expert in this field, the
concept of \emph{indeterminism}, as advocated by the
philosopher \cite{popper:open_universe}, seems to provide an
appropriate  philosophical basis for the present work.

The present method to introduce gauge fields by means of a
multi-valued dynamic variable ('phase function') has been invented
many years ago but leads, in the context of the present statistical
theory, nevertheless to several new results. In particular, it has been shown
in~section~\ref{sec:gaugecoupling}, that only the Lorentz force can exist
as fundamental macroscopic force if the statistical assumptions of
section~\ref{sec:calcwpt} are valid. It is the only force (in the absence
of spin effects, see the remarks below) that can be incorporated in a
'standard' differential equation for the dynamical variables $\rho$, $S$.
The corresponding terms in the statistical field equations, \emph{representing}
the Lorentz force, are given by the familiar gauge (minimal) coupling terms
containing the potentials. The important fact that all forces in nature follow
this 'principle of minimal coupling' is commonly explained as a consequence
of local gauge invariance. The present treatment offers an alternative explanation.

Let us use the following symbolic notation to represent the relation
between the local force and the terms representing its action in a
statistical context:
\begin{equation}
  \label{eq:IRD7SAI1NC}
\Phi,\,\vec{A} \Rightarrow
e\vec{E}+\frac{e}{c}\vec{v} \times \vec{B}
\mbox{.}
\end{equation}
The fields $\vec{E}$ and $\vec{B}$ are uniquely defined in terms of
the potentials $\phi$ and $\vec{A}$ [see~(\ref{eq:DFS32WUSS})] while
the inverse is not true. Roughly speaking, the local fields are
'derivatives' of the potentials - and the potentials are 'integrals'
of the local field; this mathematical relation reflects the physical
role of the potentials $\phi$ and $\vec{A}$ as statistical
representatives of the the local fields $\vec{E}$ and $\vec{B}$, as well
as their non-uniqueness. It might seem that the logical chain displayed
in~(\ref{eq:IRD7SAI1NC}) is already realized in the classical treatment
of a particle-field system, where potentials have to be introduced in
order to construct a Lagrangian, see e.g.~\cite{landau.lifshitz:classical}.
However, in this case, the form of the local force is not derived but
postulated. The present treatment 'explains' the form of the
Lagrangian - as a consequence of the basic assumptions listed
in section~\ref{sec:calcwpt}.

The generalization of the present theory to spin, reported in
sections~\ref{sec:spin} and~\ref{sec:spin-fish-inform}, leads to
a correspondence similar to Eq.~(\ref{eq:IRD7SAI1NC}), namely
\begin{equation}
  \label{eq:IRUD2AO9NC}
\vec{\mu}\vec{B}
\rightarrow
\vec{\mu} \cdot \frac{\partial}{\partial\vec{x} } \vec{B}
\mbox{.}
\end{equation}
The term linear in $\vec{B}$, on the l.h.s. of~(\ref{eq:IRUD2AO9NC}),
plays the role of a 'potential' for the local force on the r.h.s. The
points discussed after Eq.~(\ref{eq:IRD7SAI1NC}) apply here
as well [As a matter of fact we consider $\vec{B}$ as a unique physical
quantity; it would not be unique if it would be defined in terms of
the tensor on the r.h.s. of~(\ref{eq:IRUD2AO9NC})].  We see here a certain
analogy between gauge and spin interaction terms. Unfortunately, the
derivation of the spin force on the r.h.s. of~(\ref{eq:IRUD2AO9NC})
requires - in contrast to the Lorentz force - additional assumptions (see
the remarks in sections~\ref{sec:spin-fish-inform},~\ref{sec:spin-as-gauge}).

Our notation for potentials $\phi,\,\vec{A}$, fields $\vec{E},\,\vec{B}$,
and parameters $e,\,c$ suggests that these quantities are electrodynamical
in nature. However, this is not necessarily true. By definition,
the fields $\vec{E},\,\vec{B}$ obey four equations (the homogeneous
Maxwell equations), which means that additional conditions are required
in order to determine these six fields. The most familiar possibility
is, of course, the second pair of Maxwell's equations. A second possible
realization for the fields $\vec{E},\,\vec{B}$ is given by
the \emph{inertial} forces acting on a mass $m$ in an arbitrarily
accelerated reference frame,
see~\cite{hughes:feynmans}. The inertial gauge field may also lead to
a spin response of the ensemble; such experiments have been proposed
by~\cite{mashhoon_kaiser:inertia}. It is remarkable that the present
theory establishes a (admittedly somewhat vague) link between the two
extremely separated physical fields of inertia and QT.

It is generally assumed that the electrodynamic potentials have
a particular significance in QT which they do not have in classical physics.
Let us analyze this statement in detail. The first part of the statement,
concerning the significance of the potentials, is of course true.
The second part, asserting that in classical physics
all external influences \emph{can} be described solely in terms of
field strengths, is wrong. More precisely, it is true for classical
mechanics but not for classical physics in general. A
counterexample - a theory belonging to classical physics but
with potentials playing an indispensable role - is provided by the
classical limit~(\ref{eq:CONACL7TMF}),(\ref{eq:QHJCL74MF}) of
Schr\"odinger's equation. In this field theory the potentials play an
indispensable role because (in contrast to particle theories, like the
canonical equations)  no further derivatives of the Hamiltonian,
which could restore the fields, are to be performed. This means that
the significance of the potentials is not restricted to quantum
theory but rather holds for the whole class of \emph{statistical}
theories discussed above, which contains both quantum theory and its
classical limit theory as special cases. This result is in agreement with
the statistical interpretation of potentials proposed in section~\ref{sec:gaugecoupling}.

The precise characterization of the role of the potentials is
of particular importance for the interpretation of the Aharonov-Bohm effect.
The 'typical quantum-mechanical features' observed in  these phase
shift experiments should be identified by comparing the quantum mechanical
results not with classical mechanics but with the predictions of the
classical statistical theory~(\ref{eq:CONACL7TMF}),~(\ref{eq:QHJCL74MF}).
The predictions of two statistical theories, both of which use potentials
to describe the influence of the external field, have to be compared.

The limiting behavior of Schr\"odinger's equation as $\hbar\rightarrow 0$,
discussed in section~\ref{sec:classical-limit}, is very important for the proper
interpretation of QT. The erroneous belief (wish) that this limit
can (must) be identified with classical mechanics is closely related
to the erroneous belief that QT is able to describe the dynamics of
individual particles. In this respect QT is obviously an \emph{incomplete}
theory, as has been pointed out many times before, during the
last eighty years, see e.g.~\cite{einstein:reply},~\cite{margenau:quantum-mechanical} ,~\cite{ballentine:statistical},~\cite{held:axiomatic}. Unfortunately,
this erroneous opinion is historically grown and firmly established in
our thinking as shown by the ubiquitous use of phrases like 'the
wave function of the electron'. But it is clear that an erroneous
identification of the domain of validity of a physical theory
will automatically create all kinds of mysteries and unsolvable
problems - and this is exactly what happens. Above, we have identified
one of the (more subtle) problems of this kind, concerning the role
of potentials in QT, but many more could be found. Generalizing the
above argumentation concerning potentials, we claim that characteristic
features of QT cannot be identified by comparison with classical mechanics.
Instead, quantum theory should be compared with its classical limit, which
is in the present $3D$-case given by~(\ref{eq:CONACL7TMF}),~(\ref{eq:QHJCL74MF})
- we note in this context that several 'typical' quantum phenomena have
been explained by~\cite{kirkpatrick:quantal} in terms of classical probability
theory. One has to compare the solutions of the classical, nonlinear
equations~(\ref{eq:CONACL7TMF}),~(\ref{eq:QHJCL74MF}) with those of the
quantum mechanical, linear equations,~(\ref{eq:CONH3TMF}),~(\ref{eq:QHJH34MF}),
in order to find out which 'typical quantum-mechanical features' are already
given by statistical (nonlocal) correlations of the classical limit theory
and which features are really quantum-theoretical in nature - related
to the nonzero value of $\hbar$.
\section{Summary}
\label{sec:concludingremarks}

In the present paper it has been shown that the method reported in I, for the
derivation of Schr\"odingers's equation, can be generalized in such a way that
essentially all aspects of non-relativistic QT are taken into account. The success
of this  derivation from statistical origins is interpreted as an argument in favor
of the SI. The treatment of gauge fields and spin in
the present statistical framework led to several remarkable new insights. We
understand now why potentials (and not local fields) occur in the field equations
of QT. The non-uniqueness of the potentials and the related concept of gauge
invariance is not a mystery any more. Spin is derived as a kind of two-valuedness of
a statistical ensemble. The local forces associated with the gauge potentials,  the
Lorentz force and the force experienced by a particle with magnetic moment, can
also be derived. Apart from some open questions in the area of non-relativistic physics, a
major problem for future research is a relativistic generalization of the present
theory.

\end{document}